\documentclass{aa}  

\usepackage{graphicx}
\usepackage{natbib}
\bibpunct{(}{)}{;}{a}{}{,} 
\usepackage{hyperref}
\hypersetup{colorlinks=true, citecolor=blue}
\usepackage{color}
\usepackage{lscape}
\usepackage{longtable}
\usepackage{times}
\usepackage{nccmath}
\usepackage{float}
\usepackage{xcolor}
\usepackage{siunitx}
\usepackage{comment}

\begin{document} 

   \title{\textcolor{blue}{ Polar dust obscuration in broad-line active galaxies  from the XMM-XXL field\footnote{Table A.1 is available in electronic form
at the CDS via anonymous ftp to cdsarc.u-strasbg.fr (130.79.128.5)
or via http://cdsweb.u-strasbg.fr/cgi-bin/qcat?J/A+A/}}}
   \author{V. Buat
          \inst{1,2}
          \and G. Mountrichas \inst{3} 
          \and G. Yang \inst{4,5}
          \and M. Boquien \inst{6}
         \and  Y. Roehlly \inst{1} 
         \and D. Burgarella  \inst{1} 
         \and M. Stalevski \inst{7,8}
         \and L. Ciesla \inst{1} 
         \and P. Theul\'e \inst{1}
          }

    \institute{Aix Marseille Univ, CNRS, CNES, LAM Marseille, France
              \email{ veronique.buat@lam.fr}
              \and Institut Universitaire de France (IUF)
                 \and Instituto de Fisica de Cantabria (CSIC-Universidad de Cantabria), Avenida de los Castros, 39005 Santander, Spain    
                 \and Department of Physics and Astronomy, Texas A\&M University, College Station, TX 77843-4242, USA
                 \and George P. and Cynthia Woods Mitchell Institute for Fundamental Physics and Astronomy, Texas A\& M University, College Station, TX 77843-4242, USA
                  \and Centro de Astronom\'ia (CITEVA), Universidad de Antofagasta, Avenida Angamos 601, Antofagasta, Chile 
                  \and   Astronomical Observatory, Volgina 7, 11060 Belgrade, Serbia
                  \and    Sterrenkundig Observatorium, Universiteit Ghent, Krijgslaan 281- S9, Ghent, B-9000, Belgium            
             }



  \abstract
   {}
   {Dust is observed in the polar regions of nearby active galactic nuclei (AGN)  and  it is known to contribute substantially to  their  mid-IR emission and to the obscuration of their UV to optical emission.  We aim to carry out a  statistical test to check whether this component is a common feature  based on an analysis of the integrated spectral energy distributions of these composite sources.}
   {We selected a sample of 1275 broad-line AGN  in the XMM-XXL field, with optical to infrared photometric data. These AGN   are seen along their polar direction and we expect a maximal impact of dust located around the poles when it is present. We used X-CIGALE, which introduces a dust component  to account for obscuration along the polar directions,  modeled as a foreground screen, and  an extinction curve that is chosen as it steepens significantly at short wavelengths  or  is much grayer. By comparing the results of   different fits,  we are able to define subsamples of sources with positive statistical evidence in favor of or against polar obscuration (if present) and described using the gray or steep extinction curve.}
  {We find a similar fraction of sources with positive evidence for and against polar dust. Applying statistical corrections, we estimate that half of our sample  could contain polar dust and among them, 60$\%$  exhibit a steep extinction curve and 40$\%$  a flat extinction curve;  although these latter percentages are found to depend on the adopted extinction curves.  The obscuration in the V-band is not found to correlate with the X-ray column density, while $A_{\rm V}/N_{\rm H}$ ratios span a large range of values and higher dust temperatures are found with the flat, rather than with the steep extinction curve.  Ignoring this polar dust component in the fit of the spectral energy distribution of these composite systems leads to an overestimation of the stellar contribution. A single fit with a polar dust component described with an SMC extinction curve efficiently overcomes  this issue but it fails at identifying  all the  AGN  with polar dust obscuration.
  }
   {}
   \titlerunning{Dust obscuration  of BLAGN in the XXL field}
   \maketitle
 %

\section{Introduction}

Modeling  the characteristics of  dust   obscuration is crucial in  determining the  intrinsic spectral energy distribution (SED) of any astronomical source from the ultraviolet (UV) to the near-infrared (near-IR). It is especially crucial for quasars and active galactic nuclei (AGN)  that are strong UV emitters: even a small amount of dust can reshape dramatically their SED.
The spectral emission of AGN  and  quasars are found remarkably similar in the mid-infrared   \citep[mid-IR, e.g.,][]{Richards2006,Elvis2012,Assef2010}. The primary difference between type 1 and type 2 AGN and quasar  emission appears in the UV-optical regime and   is attributed to dust reddening \citep[e.g.,][]{Hickox2017}. On the one  hand,  the UV to near-IR emission of type 2 sources is dominated by the host component and obscured active sources are mostly identified in the mid-IR \citep[e.g.,][]{Stern2005, Donley2012, Assef2010}. On the other hand, a  large diversity of optical to near-IR SEDs  is found for AGN type 1 sources and the departure to an unobscured  type 1 SED, such as  that of \citet{Elvis1994}  is  interpreted by dust obscuration or host galaxy contamination \citep{Carleton1987, Richards2003, Prieto2010, Elvis2012, Hao2013, Yang2020}. 

The unification model \citep{Antonucci1993, Urry1995}  introduces a central dusty torus and is very successful at explaining both the bulk of mid- and far-IR emissions and the broad and narrow emission lines observed in type 1 and type 2  sources. While this model is very efficient at explaining the main differences between AGN types by the orientation of the line of sight  with respect to the dusty torus,  more complex dust configurations have also been proposed   from  parsec scales (torus) to larger scales in the host galaxy \citep[e.g.,][]{Netzer2015, Chen2015,Asmus2019, Zou2019}.

Characterizing  the nature of dust responsible of the obscuration of AGN  is very difficult and different studies have led to diverse results. For instance,  \citet{Richards2003} and \citet{Hopkins2004}  found that an SMC extinction curve gives the best fits of SEDs of quasars from SDSS and this law is commonly used to redden the intrinsic emission of an AGN \citep[e.g.,][]{Hao2013, Calistro2021}. Conversely, \citet{Maiolino2001a} found observational evidence for different dust properties in  moderately  obscured local  AGN than in the Galactic interstellar medium.  They attributed these  peculiar dust properties  to a   prominence of large grains, making the  extinction curve flatter. \citet{Gaskell2004}  derived extinction curves  by comparing UV-optical composite continuum spectra and found a significantly flatter (i.e., grayer)  curve  with no rise in the UV, which is contrary to the SMC extinction law. \citet{Czerny2004} also derived an extinction curve flatter than the SMC law by comparing blue and red composite spectra of quasars from the SDSS.    These flatter curves can be  explained with modifications of the standard dust models with    a deficit of small grains  \citep{Gaskell2004} or to a high rate of grain coagulation in a dense medium \citep{Laor1993,Maiolino2001b}.   

Thanks to mid-IR interferometry, some nearby AGN  can now be observed at parsec scales and warm dust is commonly found  distributed along the AGN polar direction \citep[e.g.,][]{Asmus2016, Lopez-gonzaga2016}. The polar dust extension may originate from dusty winds  driven by the emission of the accretion disk. Its emission represents a substantial fraction of the mid-IR emission coming from the AGN \citep{Asmus2016}. Its distribution measured on  mid-IR images is variable from tens to  hundreds parsecs; \citet{Fuller2019} report even larger scales up to one kiloparsec. 
 \citet{Stalevski2017,Stalevski2019} proposed a model to explain their high angular resolution mid-IR observations of the Circinus galaxy with a parsec scale optically thick disk and an optically thin dusty cone extending out to 40 parsecs. NGC 3783 can also be considered as a prototype  of  a  type 1 source with a  polar dust at a parsec scale composed of optically thick clouds and large carbon  grains  because of the selective destruction of silicate and small grains \citep{ Honig2017,Lyu2018}. 
  
Various dusty structures can thus contribute to the obscuration of  the nucleus emission: in addition to the equatorial dusty torus, which only affects type 2 AGN, some extra dust distributed around  the polar direction   can also redden the UV-optical emission of the accretion disk and contributes to the dust emission at IR wavelengths. The introduction of polar dust in the modeling of active galaxies of both types is found to  impact the measure of parameters  linked to dust, such  as the covering factor of obscuring material \citep{Asmus2019, Toba2021} or the fraction of dust emission coming from the AGN \citep{Mountrichas2021}.
The identification of these multiple  components and their relative contribution   is  difficult when  only integrated multi-wavelength data are available, leading to degeneracies \citep[e.g.,][]{Netzer2015, Hao2013}. Despite the  complex distribution of dust around an AGN, simplified models must be used to investigate their main features \citep{Lyu2018}.

We present  a  study of  a  sample of X-ray selected type 1 sources spectroscopically  identified  with broad emission lines (BLAGN)  in the XMM-XXL field  \citep{Pierre2016}, and with a multi-wavelength coverage \citep{Menzel2016}.  The impact  of a potential polar dust component should be easier to measure in type 1 sources. According to the unification model, the  nucleus  emission is expected to emerge from the polar direction of these objects and  we assume that any reddening of the AGN emission  comes from the polar dust. Consequently, we merge the possible large-scale effect of the interstellar medium  of the host with the reddening of a polar dust component since  we  are unable to  disentangle both components  from the integrated SEDs of the sources. For this  study, we use the X-CIGALE code\footnote{\url{https://cigale.lam.fr}
} that accounts for polar dust  modeled as a simple dust screen  in front of the accretion disk  and a dust extinction curve. We  aim to test  the presence of polar dust and   to characterize  its extinction curve. We  also study the impact of this component to the measure of physical parameters related to  host galaxy and AGN components.

In Sect. 2, we describe our  composite models (AGN and host galaxy) created with X-CIGALE to fit the SEDs of our sources and using optical-IR colors, we highlight  the impact on the optical-IR  emission of dust reddening with either a steep (SMC, \citet{Pei1992}) or a gray \citep[][hereafter, G04]{Gaskell2004} extinction curve. In Sect. 3, the SEDs of our sample of BLAGN in the XMM-XXL field are fitted with the different models presented in Sect. 2.  We define  sub-samples of sources described without any obscuration or with a polar dust and a given extinction curve in Sect. 3.  We discuss our estimations of polar dust extinction and temperature.  The  properties of the host   are compared  in Sect. 4 for the  sub-populations  with and without polar dust, and the impact of the polar dust component on the  determination of these properties is discussed in Sect. 5, while Sect. 6 is dedicated to a summary of our study.
In this paper we assume a $\Lambda$CDM cosmology with parameters coming from the seven-year data from WMAP \citep{Komatsu2011}.


\section{X-CIGALE modeling \label{sec:data}}
\begin{figure}
\centering
\includegraphics[width=8cm]{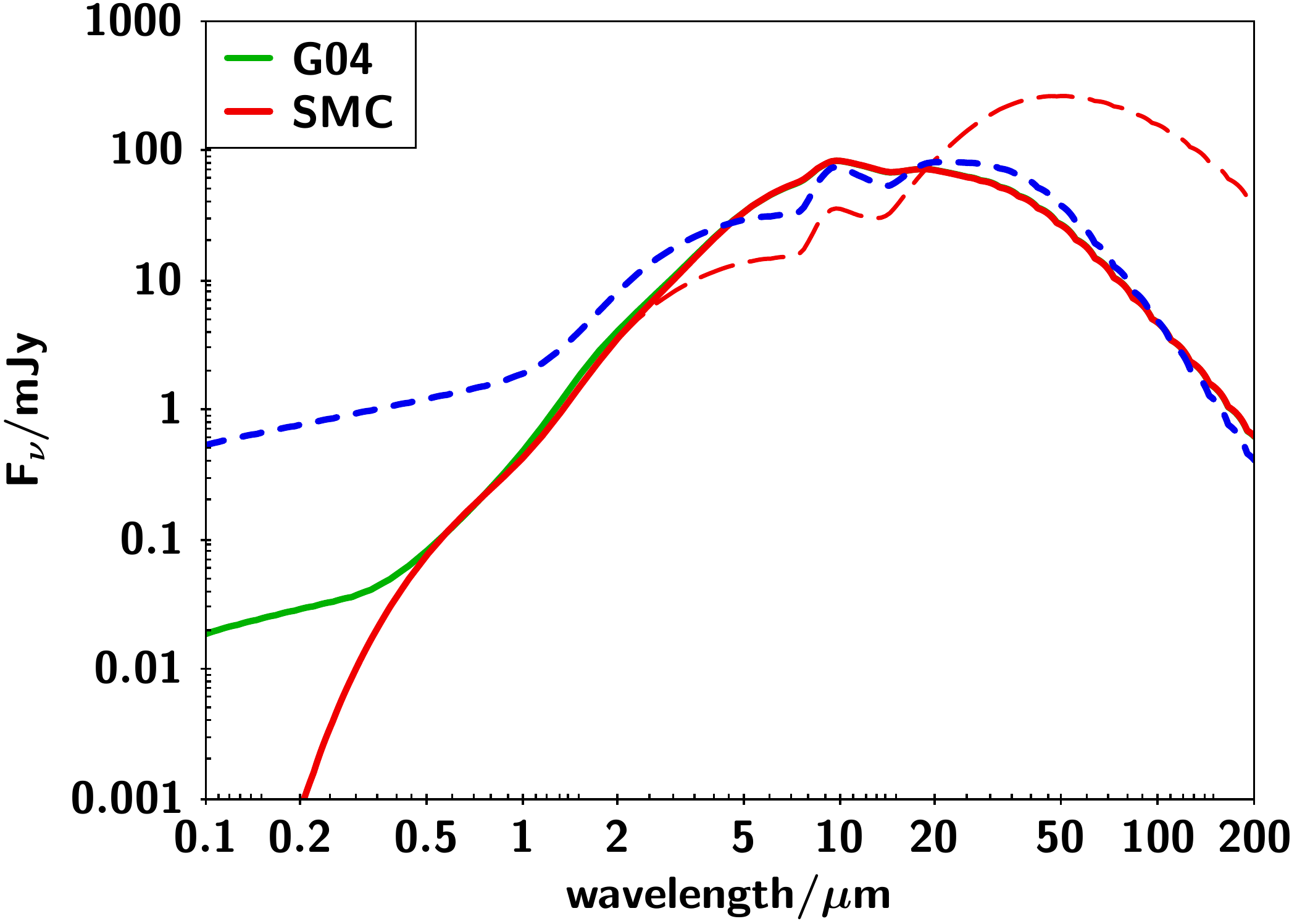}
\caption{SED of the AGN component modeled as described in Sect. 2.2.  The scale of the  fluxes  is arbitrary. The red lines correspond to a polar dust obscuration of the accretion disk emission with an SMC extinction curve, the  dust  emission is the sum of  torus   and  polar dust emission at 100 K (red solid line) and 500 K (dashed line).  The green line  corresponds to a polar dust obscuration  with a G04 extinction curve. All models are normalized to the same  energy emitted by dust. The SED of the model without polar dust is shown as a blue short-dashed line, in this case the dust emission comes only from the dusty torus and the emission of   accretion disk  is unabsorbed along the line of sight.}
\label{SEDagn}
\end{figure}

CIGALE \citep{Noll2009, Boquien2019}, along with its X-ray extension   X-CIGALE  \citep{Yang2020}, is a  very versatile code aimed at generating spectral models to fit the observed SEDs of  extragalactic systems. It  combines a stellar SED  with a dust component emitting  in the IR and conserves the energy balance between dust absorbed emission and its  thermal re-emission. An AGN component  can be added  with its emission also modeled  on the basis of the conservation of energy. The relative contribution of the AGN, $f_{\rm AGN}$,  is measured as the fraction of total dust luminosity coming from the dusty torus and   dust in polar regions (if present).  Recently, \citet{Yang2020}  developed  a new branch of CIGALE, named X-CIGALE, with a phenomenological modeling of the X-ray emission to account for X-ray  fluxes in the fits of the SED. It is this version of the code that we use in this work.


\subsection{AGN component}

\citet{Yang2020} implemented a clumpy two-phase torus model, SKIRTOR \citep{Stalevski2012,Stalevski2016},  based on the radiation transfer code, SKIRT \citep{Baes2015}.  To account  for the possible  presence  of dust in the polar regions that is not included in SKIRTOR, a new component called "polar dust" is added to the SKIRTOR model as a homogeneous dust screen in the foreground of the accretion disk characterized by  different extinction curves.  Energy conservation is also applied to this new component, and the re-emission of dust in the polar regions is modeled as a modified black body. 
In this work, we use these new developments to model a  BLAGN sample and we test the need to introduce any polar dust contribution. With our sample being X-ray selected, we ran   X-CIGALE to fit simultaneously X-ray and UV to IR data.  We considered models with and without an  extra  polar dust component described with  either  the SMC  extinction curve of  \citet{Pei1992}  or the flatter law of  G04, both implemented in X-CIGALE. The  shapes of these two curves  allow us to investigate the role of polar dust in two very different contexts \citep[Fig. 2 of ][for a comparison of extinction laws]{Lyu2014}. The G04 law expressed in $A_{\lambda}/A_{\rm V}$ is essentially flat at wavelengths lower than theV-band (Fig 3 of G04) and similar to the SMC and Milky Way extinction laws   at longer wavelengths. Conversely, the SMC law  increases steeply from the visible to  shorter wavelengths;  $R_{\rm V}$ (defined as  $R_{\rm V}= A_{\rm V}/E(B-V)$)  values of both curves also reflect their differences with  $R_{\rm V}= 2.9$ for SMC \citep{Pei1992} and $R_{\rm V}= 5.5 $ for the analytical formula of the G04 curve implemented into X-CIGALE.

To illustrate the impact of  the choice of the  extinction law, in Fig. \ref{SEDagn}, we show the SED of the AGN component without polar dust and with a polar dust extinction calculated with either an SMC or a G04 law. The modification of the disk emission   is  clearly visible at rest-frame wavelengths lower than 0.5 $\mu$m. With the SMC law the intensity of the disk emission decreases sharply with wavelength, whereas it  keeps its original shape  when the  G04 law is applied.

We choose to compare SMC and G04 extinction curves as a first attempt to discriminate between two very different shapes of the AGN SEDs. The G04 curve was obtained for radio-loud AGN and is flatter than the extinction curve derived by \citet{Czerny2004} with composite spectra of red quasars from the Sloan Digital Sky Survey (SDSS). From their study of individual AGN,  \citet{Gaskell2007} found a large diversity of extinction curves and derived an average extinction curve that is not as flat as the G04 curve. In order to test the impact of the adopted extinction curves,  we also considered the average extinction law of \citet{Gaskell2007} in our analysis.

\subsection{Composite models, AGN, and host galaxy}

\begin{figure}
\includegraphics[width=8cm]{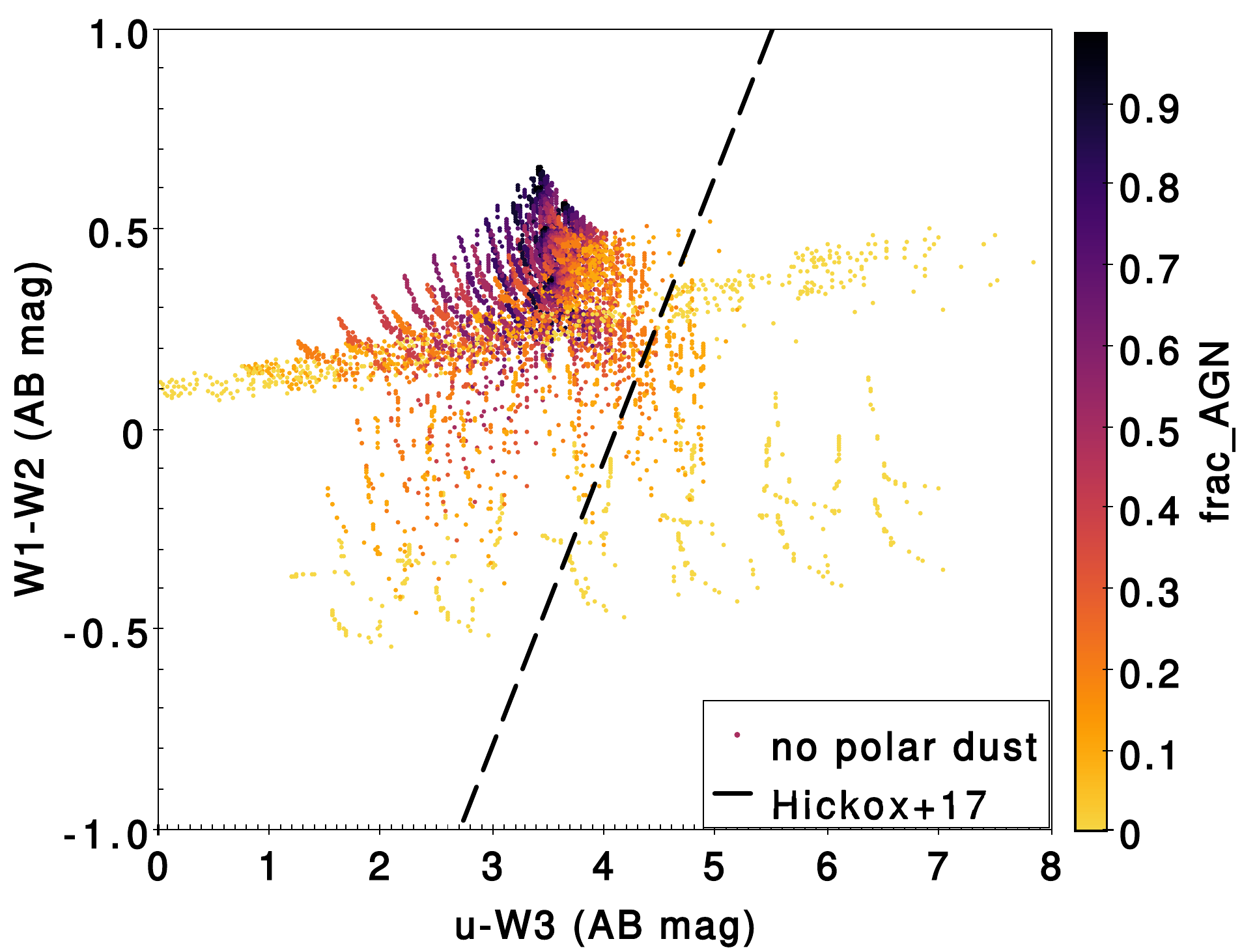}
\includegraphics[width=7cm]{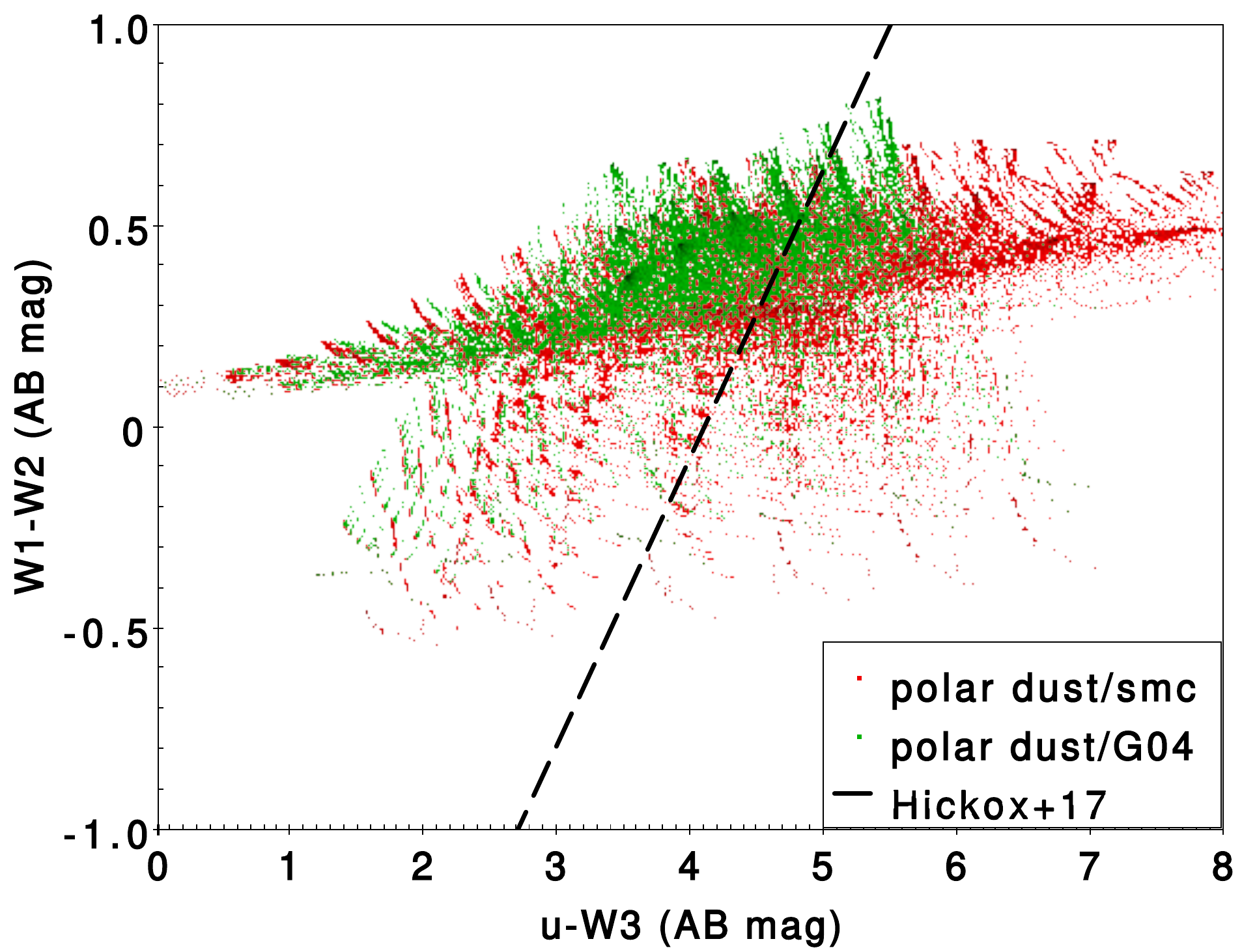}
\caption{Distribution of composite models in the {\it u}-W3, W1-W2 color plot. The models are created with X-CIGALE for an AGN type 1 and a host galaxy from $z=0$ to $z=2.5$. {\it Top panel:} No  obscuration by polar dust is considered. The relative contribution of both components defined  by the fraction of the dust luminosity coming from the AGN is color coded. {\it Bottom panel:}  Polar dust obscuration  is added. Models generated with the SMC (resp., G04) extinction curve  are plotted as red (resp., green). The SMC extinction curve leads to a larger reddening of the {\it u}-W3 color.  All magnitudes are in AB units and the line boundary defined by \citet{Hickox2017} to identify obscured quasars  is modified accordingly ($u-\rm{W3}=1.4 \times(\rm{W1-W2})+4.1$).The number of G04 models is reduced by a factor of 2 on the figure for a better visibility for both distributions.}
\label{models}
\end{figure} 

To highlight the effect of the polar dust component, we  consider all our  composite models built with X-CIGALE prior to the fitting process. They are composed of  an  AGN type 1 template  and of a stellar component.  

We use a delayed star formation history (SFH) with the functional form  ${\rm SFR} \propto t \exp(-t/\tau_{\rm main})$ for  the star formation rate (SFR) of  the main stellar population. A recent constant burst of star formation is over-imposed whose amplitude is measured by the stellar mass fraction produced during the burst,  $f_{\rm burst}$ \citep[e.g.,][]{Buat2018}. The ages of the two stellar populations  are flexible. We adopt  the Initial Mass Function  of \citet{Chabrier2003} and the stellar models of \citet{bc03} with a metallicity  fixed to the solar value.
The stellar emission is attenuated with the   law of \citet{Calzetti2000}, and the absorbed energy is re-emitted   using \citet{Dale2014}\footnote{The  \citet{Dale2014} templates are used without  AGN contribution which is added with the SKIRTOR module} templates parametrized by a single parameter $\alpha$  representing the relative contribution of different local  heating conditions.  

Following \citet{Yang2020}, the AGN type 1 component is modeled by fixing  the viewing angle  to 30 degrees. All the other fixed  input parameters describing the torus model of SKIRTOR are  fixed   as in \citet{Yang2020} (their Table 2). We use the disk continuum emission from SKIRTOR with four power-laws to cover the full range from 0.008 $\mu$m to 1000 $\mu$m.  In a new version  of X-CIGALE, Yang et al.  (2021, in preparation) introduce a flexible power-law indice for $0.1<\lambda<5 {\rm \mu m}$  as $\lambda F_\lambda \propto \lambda^{-0.5+\delta_{\rm AGN}}$  to better reproduce the colors of SDSS quasars, $\delta_{\rm AGN}=0$ corresponding to the original SKIRTOR model. The determination of this parameter  is difficult because of its degeneracy with the amount of obscuration generated by the polar dust component.  To overcome this issue, we tested  a variable $\delta_{\rm AGN}$ on our sub-sample defined without polar dust (Sect. 3.3.1). We found an average (median) value of -0.12 (-0.2), but the parameter was not  securely determined. We then  tested its influence on our analysis by fixing $\delta_{\rm AGN}=-0.2$,  the percentages of sources found in each category (with or without polar dust, c.f. Sect. 3)  were only slightly modified (at most $10\%$). We consider this effect small as compared to all the other sources of uncertainty in our analysis and we keep the original continuum emission of the SKIRTOR model.

Three sets of models were built: without any obscuration applied to  the AGN emission or with a polar dust extinction modeled as a screen and  an SMC or a G04 extinction curve. The amount of dust absorption is quantified with  the color excess $E(B-V)_{\rm pd}$. Polar dust re-emission is modeled with a modified black-body with $\beta=1.6$ and the wavelength where the optical depth is equal to unity is equal to 200 $\mu$m \citep{Casey2012}. The temperature of the polar dust is   allowed to vary from 100 to 2000 K to roughly represent the different scales of polar dust exposed to the radiation of the accretion disk \citep{Lyu2018,Stalevski2019}.  The input values of the free parameters  used to build the models  are presented in Table \ref{param}.

\begin{table}
\scriptsize
\centering
\begin{tabular}{l c c}
\hline
 \multicolumn{3}{c}{Stellar component}\\
\hline
\multicolumn{3}{c}{Star Formation History and dust attenuation: }\\
\multicolumn{3}{c}{delayed modeled with a recent burst }\\
\hline
age of the main population & $\rm age_{main}$ & 1.5, 2, 3, 4, 5, 7  Gyr \\
 $e$-folding time&$\rm \tau_{main}$& 0.5,1,2,3,4,5 Gyr \\
age of the burst & $\rm age_{burst}$  &100, 500 Myr\\
burst stellar mass fraction & $f_{\rm burst}$ & 0.0, 0.01, 0.02, 0.05, 0.1, 0.15, 0.2\\
\hline
\multicolumn{3}{c}{Dust absorption \& re-emission}\\
\hline
 Calzetti 2000 law, &$E(B-V)$   & 0.05, 0.1, 0.2, 0.3, 0.4, 0.5 \\

 Dale et al. IR templates &$\alpha$ slope   & 1.5, 2.0, 2.5\\
\hline
\multicolumn{3}{c}{AGN component: SKIRTOR2016}\\
\hline
viewing angle && 30 deg\\
AGN  dust luminosity fraction &$f_{\rm AGN}$&0,0 to 0.9 (0.1 step)\\
Polar dust extinction law & & SMC, Gaskell2004  (G04)\\
E(B-V) of polar dust &$E(B-V)_{\rm pd}$& 0.0,0.01, 0.1, 0.15, 0.2, 0.25, 0.3\\
Temperature of Polar dust &$T_{\rm pd}$ & 100, 500, 1000, 2000 K\\
\hline\hline
\end{tabular}
\caption{X-CIGALE modules and input variable parameters used to create the stellar and AGN components of the composite models. We refer to \citet{Yang2020} for the description and values of  the input fixed parameters which are not described here. See text for details.}
\label{param}
\end{table}

In Fig. \ref{models}, the full set of models is represented in an optical-IR color   diagram combining photometric bands from the SDSS ({\it u}) and {\it Wide-field Infrared Survey Explorer} ({\it WISE})  (W1, W2, and W3) surveys. The impact of dust extinction affecting the UV-optical continuum of the AGN  is clearly visible in this {\it u}-W3, W1-W2)  color diagram  and   \citet{Hickox2017} proposed that obscured and unobscured  quasars  can be effectively  distinguished  with obscured sources corresponding to $\rm {\it u}-W3 > 1.4 (W1-W2)+4.1$ (in AB units). 
Models without polar dust are plotted in the top panel. They span a limited range of colors as also shown by \citet{Hickox2017}. By comparison the locus of models with polar dust (bottom panel of Fig. \ref{models}) is extended to larger  $\rm {\it u}-W3$ colors. The additional obscuration reduces  the flux in the {\it u} band. The reddening of  $\rm {\it u}-W3$ is stronger with the SMC extinction curve, which steeply  increases at short wavelengths than with  the much flatter G04 curve.

\section{Analysis of  BLAGN  from the XMM-{\it XXL} survey}

\subsection{Sample}

The X-ray AGN sample used in this work comes from the XMM-XXL survey \citep{Pierre2016}.  This survey has a medium depth of $\rm \sim 6 \times 10^{-15}\,erg\,cm^{-2}$\,s$^{-1}$ and an exposure time of $\sim$10\,ks per XMM pointing. The field covers a total area of 50\,deg$^2$ split into two nearly equal extragalactic sub-regions. In this study, we use the XXL North sample that consists of 8445 X-ray sources \citep{Liu2016}.  Spectroscopic redshifts and spectral classification  are given by \citet{Menzel2016} for 2512 of these sources. In our analysis, we start with the 1637 BLAGN of the Menzel et al. catalog, with a spectroscopic redshift lower than 2.5. Intrinsic X-ray fluxes are taken from \citet{Mountrichas2021}.

The available photometry is described in Sect. 2 of  \citet{Mountrichas2021}. In summary,  all our sources have  {\it u,g,r,i,z} photometry from the Sloan Digital Sky Survey (SDSS). Near-IR  photometry (J,H,K)   is available for 836  sources  from the Visible and Infrared Survey Telescope for Astronomy  (VISTA; \citet{Emerson2006}. Mid-IR photometry  comes from   {\it Spitzer} (IRAC and MIPS)  and  all{\it WISE} \citep{Wright2010} datasets. {\it Spitzer} data are   provided  by the HELP collaboration \citep{Shirley2021}. The wavelength coverage  of {\it WISE} and IRAC is crucial to identify AGN and we reduce the sample to the 1523 sources detected with either IRAC1 or W1, and IRAC2 or W2, by order of priority.  We  ensure a mid-IR coverage  by selecting  sources with at least one measured flux  in either W3, W4, or MIPS1\footnote{To consider only {\it WISE}  detections we discarded all fluxes with no associated uncertainty}.We are left with 1275 sources. {\it Herschel}/SPIRE fluxes for 824 of these sources come from the HELP project\footnote{The {\it Herschel} Extragalactic Legacy Project (HELP) is a European funded project to analyze all the cosmological fields observed with the {\it Herschel} satellite. The HELP datasets are available on the Herschel database in Marseille  (\url{http://hedam.lam.fr/HELP/}).}.  Only  SPIRE fluxes were included in the analysis because of  the much lower sensitivity of the PACS observation in the  XMM-XXL field \citep{Oliver2012}.
\begin{figure}
\centering
\includegraphics[width=8cm]{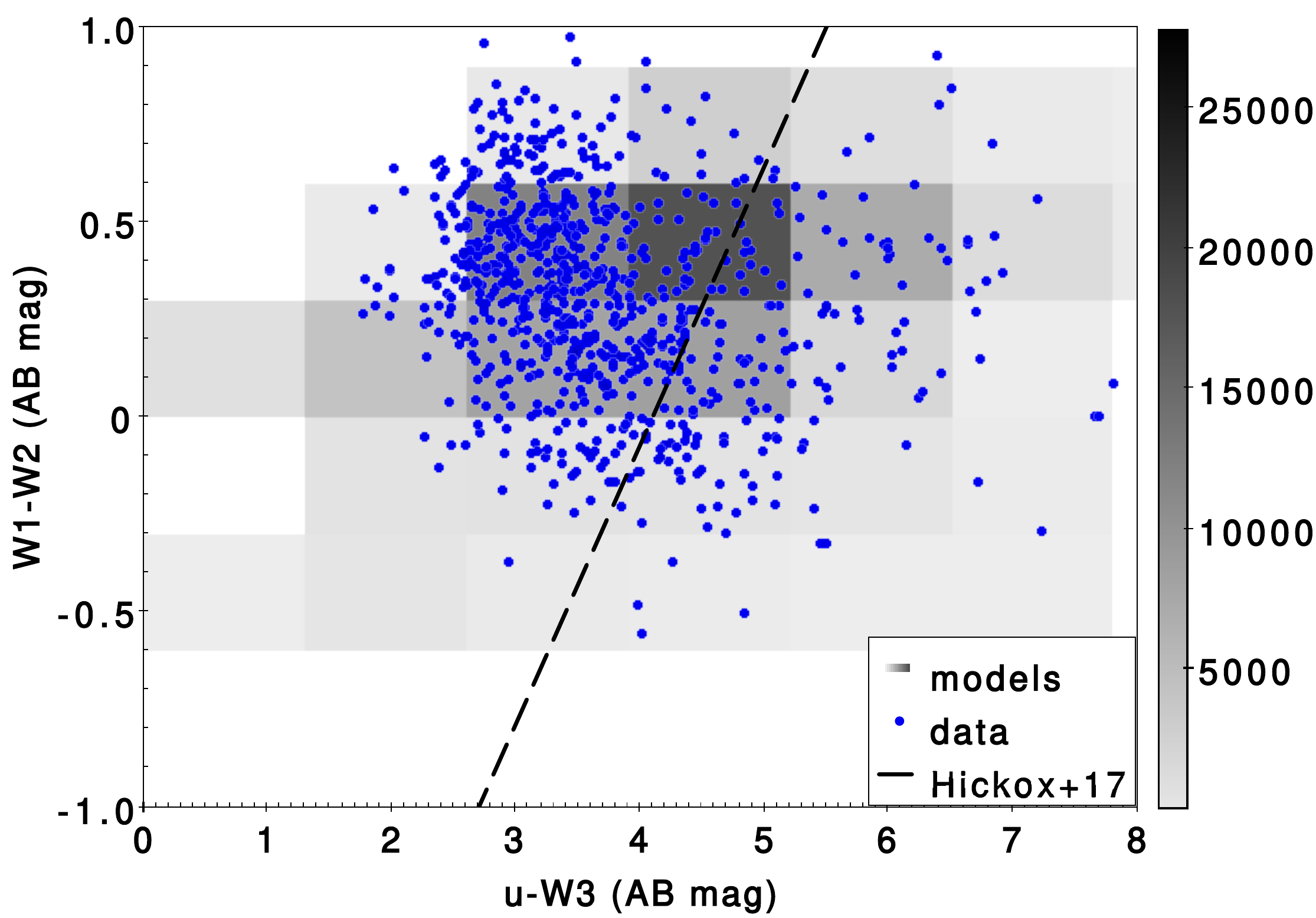}
\caption{Models compared to data in the $\rm {\it u}-W3, W1-W2$ color diagram.  The density of models created with X-CIGALE is plotted with a gray scale and the observed colors in blue. $\rm W1-W2$ is calculated with IRAC or {\it WISE} fluxes, by order of preference.}
\label{allmodels}
\end{figure} 

\subsection{SED fitting and SFR and $\rm M_{star}$ estimations}
In Fig. \ref{allmodels}, we compare the color distributions of models and data  in the same  color plot as was previously used to show the models in Fig. \ref{models}. To plot the W1-W2 color, we use either IRAC or {\it WISE} data without applying any correction  which are less than 0.2 mag \citep[e.g.,][]{Richards2015}. 
The set of models is found to cover  the observed colors.  The need to introduce a reddening of the AGN emission for at least a fraction of the sources is clearly set out when Fig. \ref{allmodels} is compared to the distributions of models  in Fig. \ref{models}.

Three different fits are performed  with the  models   described in Sect. 2: without  polar dust and with a polar dust component described by  an SMC  extinction curve and then a G04 . The global quality of the fit is assessed by  
the $\chi^2$ value of the best fit and by  a reduced  $\chi_r^2$, defined as the   $\chi^2/(N-1)$  with $N$ being the number of data points\footnote{$\chi_r^2$  value is only used as an indication of the global quality of the fit and the best model is not used for the estimation of the parameters}. The values of input and output  parameters and their  corresponding uncertainties are estimated as the  likelihood weighted means and standard deviations. To exclude badly fitted SEDs, we only consider sources for which at least one of the three fits correspond to $\chi_r^2<5$; thus, 1212 ($95\%$) sources fulfill this condition. Our conclusions are unchanged, with a limit set at $\chi_r^2<3$ (1030 sources).

In the next sub-section, we compare the  three fits using the Bayesian  information criterium (BIC), defined as ${\rm BIC}= \chi^2+k \times \ln(N),$ where   $k$ is the number of free parameters \citep{Ciesla2018, Buat2019}. BIC scores a model on its likelihood and complexity:  a penalty is put on the model with the highest number of free parameters and the model with the lowest BIC is prefered. When  polar dust obscuration is considered with a fixed extinction curve, two free parameters are added (dust temperature and color excess), as compared to the fit without polar dust.  

Our sample spans  a large range of redshifts, and performing a single modeling of the SFH   (cf. Table \ref{param}) would not correctly describe  the bulk of the stellar content. In order to test the validity of our single fit, we tested another set of parameters for the   SFH.  We defined five redshift bins from z=0 to 2.5 and  in each bin we defined an age for the main stellar population in each bin close to the age of the Universe. A recent burst was added in a similar way as for the first fit. The results of our study were found unchanged  with regard to the  need to account for polar dust and for the best extinction curve to model it. The SFR and stellar mass ($\rm M_{star}$)  estimates were  also found to be similar. For the sake of simplicity, we kept our single model of SFH with a large set of input parameters.
 
 Using mock data generated with X-CIGALE, \citet{Mountrichas2021} checked  the  robustness of measurements of SFR and  $\rm M_{star}$ measurements for samples of AGN galaxies with similar redshift range and  wavelength coverage. These authors also tested the effect of considering or not  SPIRE data on SFR measurements. We performed all these checks for our sub-sample of BLAGN and reached similar conclusions, attesting the robustness of SFR and  $\rm M_{star}$ determinations.  
To further exclude unreliable estimations of SFR and  $\rm M_{star}$, we applied  the same method  as \citet{Mountrichas2021}  and compared the value of the best model with the likelihood weighted mean value calculated by X-CIGALE. A large difference between them means that the probability density function is either ill-defined or very asymmetric, in both cases the estimation of the corresponding parameter is not satisfying. Therefore, in the following, we consider only measurements corresponding to $\lvert\log(\rm SFR^{best}/SFR^{bayes})\rvert < 0.5$ dex and  $\lvert \log(\rm M_{star}^{best}/M_{star}^{bayes})\rvert< 0.5 $ dex.
\begin{figure}
\centering
\includegraphics[width=8cm]{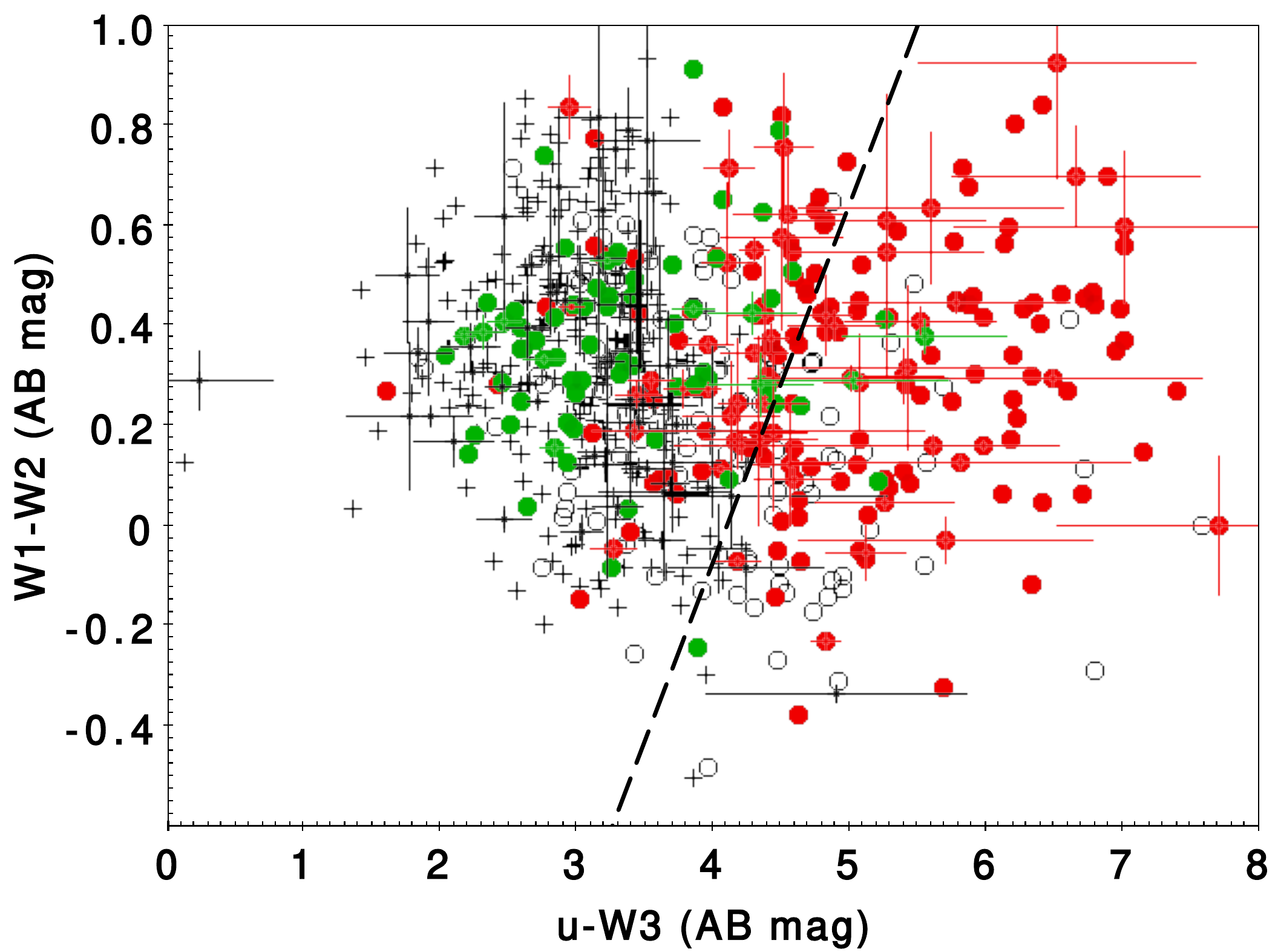}
\caption{Distribution of our BLAGN sample in the same color plot as in Fig. \ref{models}. Sources for which a polar dust component  is not needed are represented with crosses, sources better fitted with  a polar dust component are plotted with circles. Green circles are for sources better fitted with a G04 extinction curve, red circles for sources better fitted with an SMC extinction curve, open circles sources for which the extinction curve cannot be determined. Sources populating the upper right side of the plot are predominantly found with a polar dust characterized by the SMC extinction curve, the reddening of the $\rm {\it u}-W3$ color is strongly reduced with the G04 extinction curve. Magnitudes are in AB units and the dotted line represents the boundary defined by \citet{Hickox2017}  and modified accordingly ($\rm {\it u}-W3=1.4 (W1-W2)+4.1$). Uncertainties in the measured colors are over-plotted for one source randomly selected over five to avoid crowding.}
\label{extcurve}
\end{figure}

\subsection{ Polar dust  component}
We focus now on the study of  polar dust, by first defining sources for which  this component  is needed (or not) to substantially improve  the fit and, in a second step,  determining which extinction curve is preferred.

\subsubsection{ Selection of sources according to the polar dust component}
To compare our  fits and select a model, we calculate the difference of BIC between  pairs of models (SMC and no polar dust; G04 and no polar dust; SMC and G04)  for our 1212 sources. We adopt the criterium of positive evidence against the model with the highest BIC when $\Delta(BIC)>2 $  (positive evidence) \citep{Kass1995}.  
We find a positive evidence against no polar dust  for  347 objects (29$\%$). 
We also select  sources for which no polar dust is needed to fit their SED. The models we are comparing   are nested:  models  with polar dust  and $E(B-V)_{\rm pd}=0$ correspond to models without polar dust. Thus, we   consider that no polar dust is needed  when both best values  of $E(B-V)_{\rm pd}$ with either SMC or G04  are found equal to 0 \citep[e.g.,][]{Aufort2020}, it is the case for  332 sources ($27\%$). Nothing can be concluded for $44\%$ of the sample.

Among the  347  sources that are better described with polar dust, an SMC extinction curve is preferred (as a positive evidence against G04) for $50\%$ of them (173 sources). Only $21\%$ (72 sources) correspond to  a G04 curve and  there is no significant difference between the two extinction curves for  the remaining 29$\%$.  

We compare these numbers to   those found when only sources with  SPIRE data are considered. With this restricted sample of 790 sources,   no polar dust   is needed  for $22\%$ of the  sources, while $31\%$  show positive evidence against no polar dust with a   repartition between either  SMC (44$\%$) and G04 (26$\%$) extinction curves. These percentages are  close the ones we find with the full sample. 

In the following, for the sake of clarity, we define  a  BLAGN  with polar dust as one of the 347 sources  {\bf with a positive evidence against   no polar dust}; 173  will be considered as described with an SMC law and 72  with a G04 law (i.e., only 71\% of BLAGN with polar dust will have an assigned extinction law). The subsample defined as BLAGN without polar dust will consist of the 332 sources  best fitted  with $E(B-V)_{\rm pd}=0$ for each polar dust model. In  Fig. \ref{extcurve}, these different subsamples are plotted in the  {\it u}-W3, W1-W2 plane. It can be seen that most of the BLAGN sources populating the upper right region are found with a polar dust and an SMC extinction curve: a strong reddening is needed to shift the {\it u}-W3 color of these sources from the locus without polar dust to the observed values. Sources that are better characterized by the presence of polar dust and the much flatter G04 extinction curve are predominantly found in the left region of Fig. \ref{extcurve}, where models without polar dust also lie: the color reddening is strongly reduced  with the G04 law as shown in Fig. \ref{models}. 
 The status of the polar dust component for each of the 1212 sources is listed in Table A.1, the full version of the table is given online.

\subsubsection{Statistical corrections from a mock analysis}

In order to further check the validity of our results on the introduction of the polar dust component, we  performed a specific mock analysis with the full dataset.  We used the X-CIGALE option to create a mock dataset from the best SED of each fit  for  the  three subsamples of sources without polar dust, with polar dust modeled with the SMC law,  and with polar dust modeled with the G04 law. The code uses the best flux of each source and adds a random noise  extracted from a Gaussian distribution with the same standard deviation as for the observed flux. We merged the three datasets  in a single mock  catalog. This simulated  catalog  is  fitted with the same strategy as for the observed sample (cf. Sect. 3.2).  We performed the  selection of sources as described  in Sect. 3.3.1:  we chose the mock sources with    positive evidence against no polar dust  and described either by SMC or G04 extinction law, as well as  the sources whose best fit of their SED corresponds to  no polar dust. 

 52$\%$ of the input   catalog created without polar dust is confirmed with no polar dust. The percentage of successful identifications of   input sources with polar dust is   60$\%$.  Nothing can be concluded for 48\% (resp. 40\%) of sources built without polar dust (resp. with polar dust). The number  of misidentifications is very low, with less than 2$\%$ of mock sources with polar dust classified without polar dust and no mock source without polar dust classified with polar dust. From this mock analysis,  we  conclude that the fraction of observed BLAGN  classified with and without polar dust are likely to be similar since the  fractions measured are similar (29$\%$ and 27$\%,$ respectively) and the corrections to apply to both populations  close to each other (respectively 60$\%$ and 52$\%$ of   sources recovered). Applying these corrections to the observed fractions leads to $\simeq 50\%$ of sources either with or without polar dust.

The  distinction  between SMC and G04 extinction curves is also investigated with our simulated sample. The percentage of mock sources created with the G04 law and  identified  successfully is $44\%$ , the percentage reaches   $67\%$ for the SMC law.  If we account for the full process  of  selection (evidence for or against polar dust and then distinction between G04 and SMC laws), the identification of the true extinction law is  successful for 49$\%$ of the input sources with the SMC curve against 31$\%$ for the G04 curve. This analysis  confirms  the difficulty in identifying polar dust characterized by a gray extinction curve. If we apply the corrections drawn from our mock analysis  to correct the numbers of  sources observed  with  SMC  o G04 law (173 and 72, respectively), we find  that 60$\%$  of the sources with polar dust should be characterized by a steep (SMC) law   and 40$\%$ with a flat (G04) law.


\subsubsection{Impact of  the extinction curve}
The G04 law can be considered as an extreme, it was derived for radio-loud AGN and is flatter than other curves derived on different AGN and quasar populations \citep[e.g.,][]{Czerny2004}. In this section, we investigate the impact of the adopted extinction curves by adopting   the average extinction curve proposed by  \citet{Gaskell2007} to supersede the G04 law (hereafter, the GB07 law). 

 We performed the same analysis as in Sect. 3.3.1, this time with the SMC and GB07 laws.
The percentage of sources whose SED is better described with or without polar dust are only slightly modified: it increases from  27\% to 33\%  for sources without polar dust and is found nearly unchanged for sources with polar dust  (28\% to 27\%).
 The slight increase of sources for which no polar dust is needed comes from the way we select these sources,  when the  best values  $E(B-V)_{\rm pd}$ of both fits  with polar dust is equal to 0 (cf. Sect. 3.3.1). Adopting the  GB07 law  leads to more reddening than with the G04 law for a given $E(B-V)_{\rm pd}$:  some  sources for which only a very small   reddening is acceptable  can be  best fitted with G04 and a low  value $E(B-V)_{\rm pd}$; but not with GB07, for which  the best fit corresponds to   $E(B-V)_{\rm pd}=0$. 

As expected, the assignment of a dust extinction curve, either steep (SMC) or  flat (G04 and GB07), is found to depend on the curves considered for the fits.
When we use  the  SMC and GB07 laws,  43\% (134 sources) of sources with polar dust are described as SMC against 50\% (173 sources)  with  the SMC and G04  laws. No extinction curve can be assigned to the 39 sources described with SMC polar dust in the SMC/G04 fits  and not in the  SMC/GB07 fit. This result is  explained by the smaller difference between  the shape of the SMC and GB07 curves  than between SMC and G04.  For these 39 sources, the best $\chi^2$  obtained with GB07 are  found lower than with G04. Their    polar dust component may be characterized by an extinction curve that is intermediate between SMC and GB07. 

The major change  we find  is  for the assignment of the flat law: this is the case for 21\% (72 sources) with  G04  but for  only 7\% (24 sources)  with  GB07. To understand this difference, we analyzed  the 72 sources previously described  with  G04. We could confirm only 47 of them  as having polar dust with the SMC/GB07 fit, nothing can be concluded with regard to the presence of polar dust for the remaining  25.
Among the 47 sources with polar dust, 20 are found  better described  with  GB07 polar dust,  however, no conclusion can be  reached for the remaining 27. We conclude that the strong decrease in the number of sources with polar dust described with  GB07 is explained by the  larger fraction of fits for which  no significant difference is found between models with no polar dust  and  with polar dust, either SMC or GB07.
The best  $\chi^2$ found for the fits of the 72 SEDs  with GB07 are systematically higher   than with G04.  
The sources identified with  G04 polar dust  do not exhibit any substantial reddening, as shown in Fig. \ref{extcurve}, and their UV-to-far IR  SED is better fitted with a very flat extinction curve. 

Obviously, our very simplified comparison of only two laws cannot reflect the large  diversity of  AGN  dust extinction laws reported  both at low and high redshift  \citep{Crenshaw2001, Maiolino2001b, Gaskell2007, Gallerani2010, Gaskell2016}. It is worth noting that such a diversity is also found for the effective attenuation curves  for normal galaxies \citep[][and references therein]{Salim2020}. In our analysis,   we cannot separate the effect of polar dust close to the AGN  or polar dust that is distributed on larger scales in the host galaxy.  In a future work, we plan to implement flexible extinction recipes to study the  diversity of  polar dust  extinction curves and investigate correlations with other characteristics of AGN and host galaxies.

\subsubsection{Properties of  polar dust}
 \begin{figure}
 \centering
 \includegraphics[width=8cm]{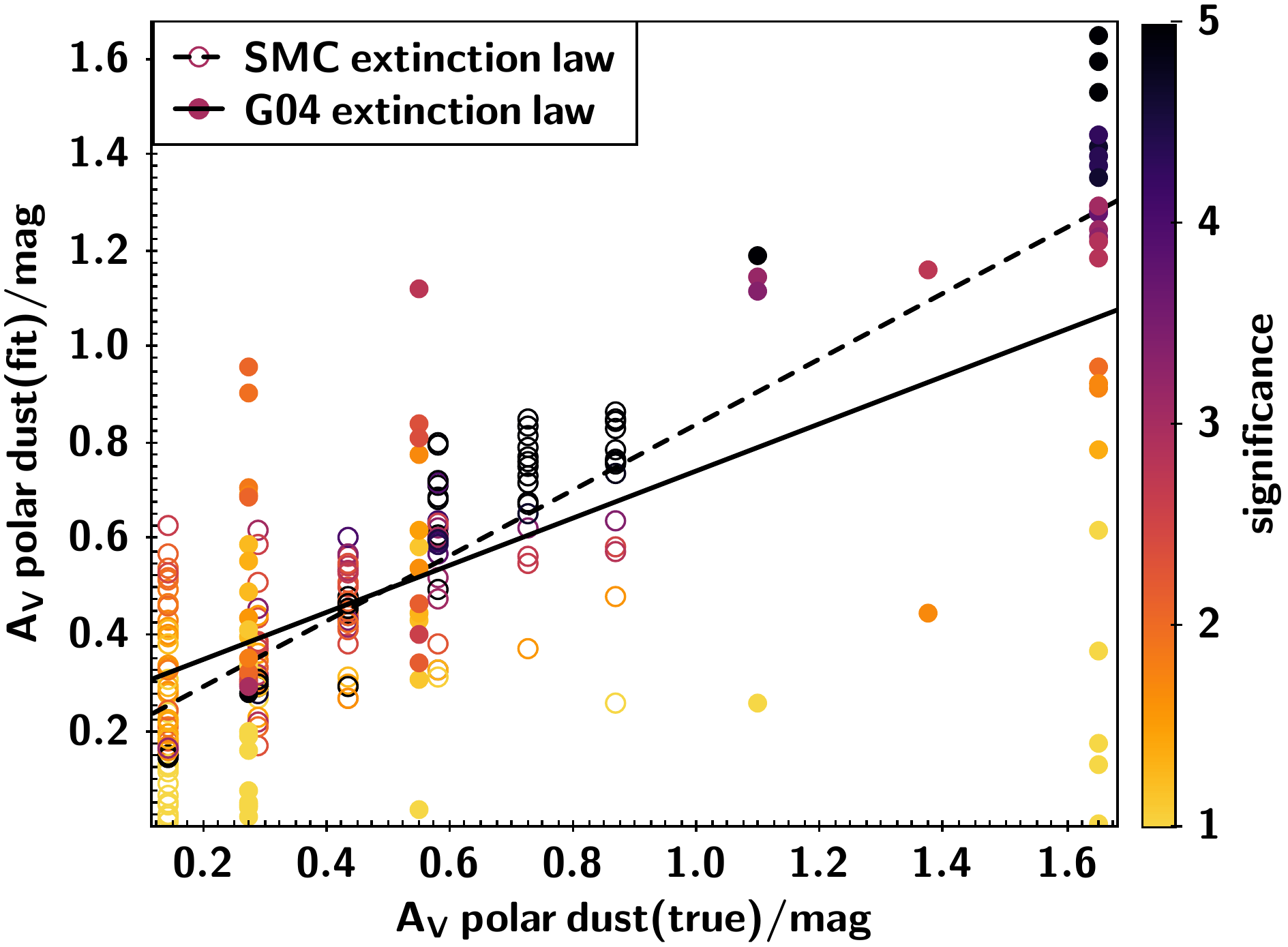}
\includegraphics[width=8cm]{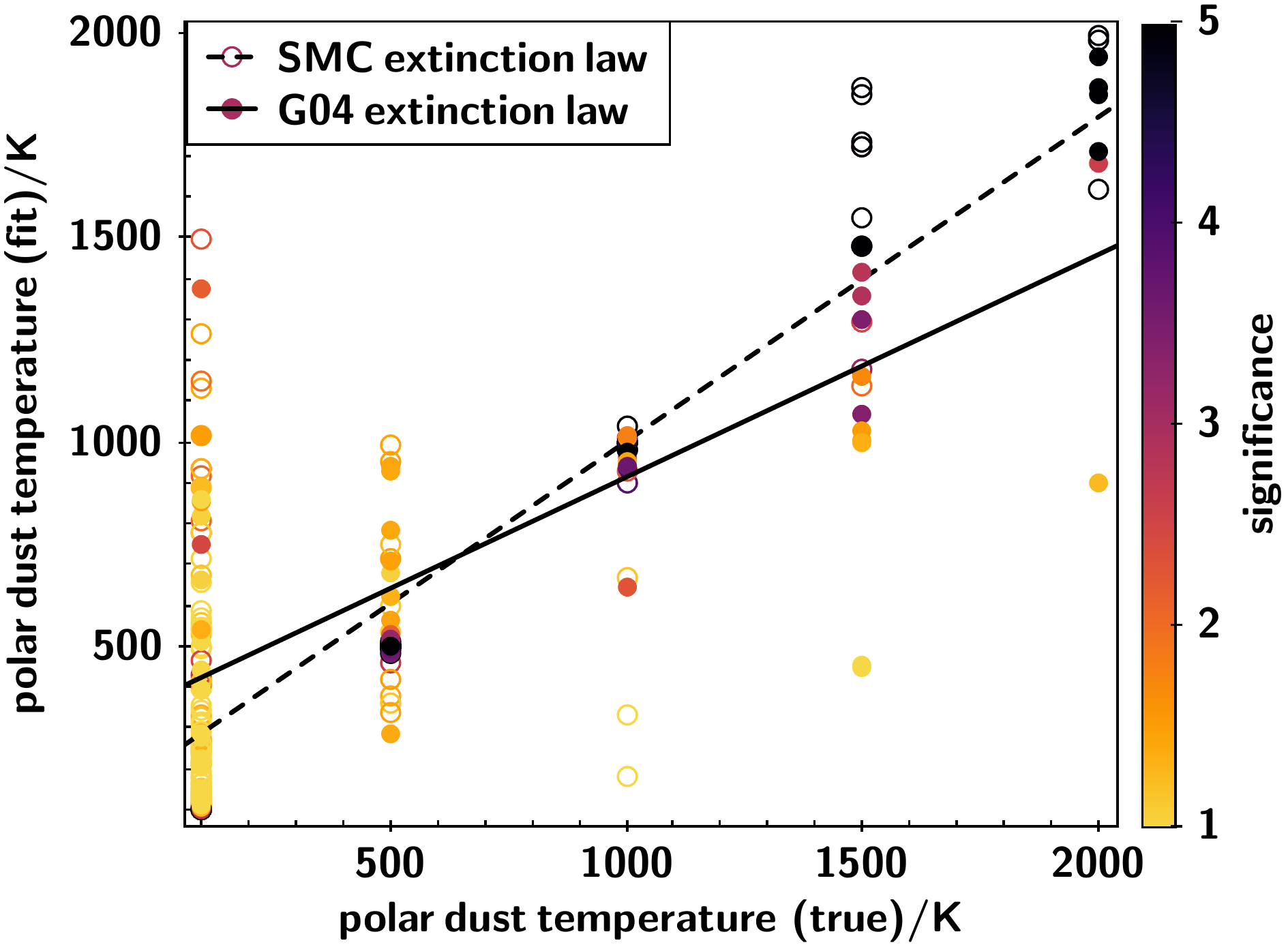}
\caption{Results of the mock analysis performed for polar dust extinction in the V-band, $A_{\rm V}$ ({\it top panel}) and polar dust temperature ({\it bottom panel}). The exact values (x-axis) are plotted against the values measured   by fitting the simulated datasets (y-axis). The open circles refer to the SMC extinction curve, the filled circles to the G04 extinction curve. The linear fits are shown as solid (G04) and dotted (SMC) lines. The statistical significance of the measure is color-coded.}
\label{mockT}
\end{figure} 

\begin{figure}
 \centering
\includegraphics[width=8cm]{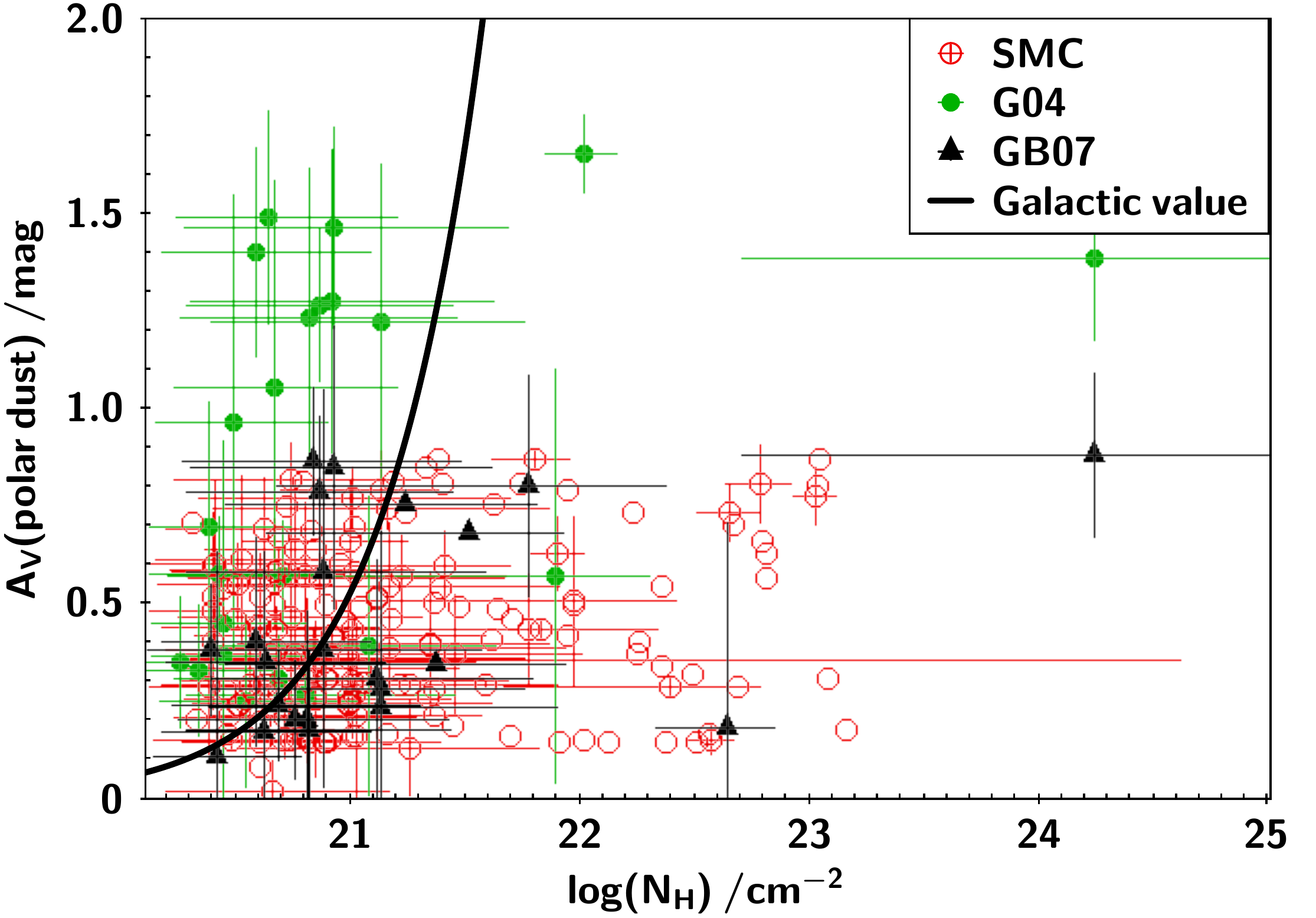}
\includegraphics[width=8cm]{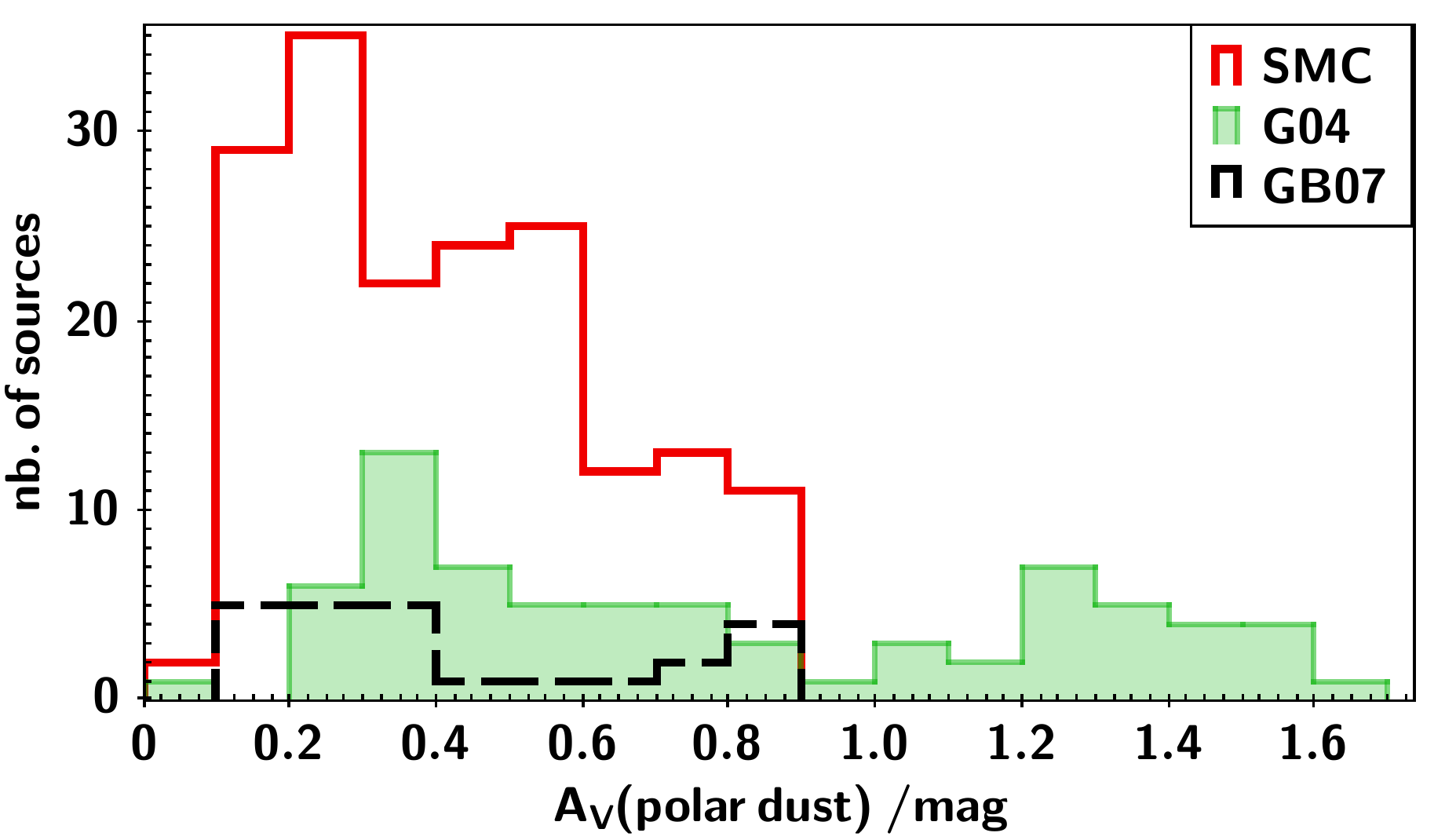}
\caption{V-band attenuation $A_{\rm V}$ due to  the AGN  polar dust component. {\it Top panel}: $A_{\rm V}$ is plotted against the X-ray absorbing column density  $N_{\rm H}$ for sources with  polar dust described either with the    SMC law (red circles) or the G04 law (green points).  {\bf $A_{\rm V}$}  is also plotted   for  the 24 sources that are found to be consistent with the GB07 law (black triangles). Uncertainties on $A_{\rm V}$ measurements (1$\sigma$ dispersion estimated by X-CIGALE) and  $N_{\rm H}$ (16th and 84th percentiles of the distribution) are over-plotted (for only one source  with SMC law  randomly selected over 3).  The Galactic value of $A_{\rm V}/N_{\rm H}$  is indicated as a  black line.  {\it Bottom panel}: $A_{\rm V}$ distribution for sources described with each law.}
\label{AV}
\end{figure} 

In this part of the study, we go back to our initial fits and results with the SMC and G04 laws and discuss the impact of using the GB07 law.
Prior to the analysis of some  polar dust  properties inferred from our analysis, we recall that we are  modeling  the component that we call "polar dust,"  with  a dust screen along the direction perpendicular to the plane of the dusty torus and an extinction calculated with a single extinction curve. This very simple model does not allow  us to get any direct information on the local and large-scale distribution of this dusty component distributed in the polar direction.

  Here, we  discuss two output quantities of our analysis: the amount of extinction in the V-band  due to polar dust and the temperature of the polar dust. The estimated values of these two parameters are given  in Table A.1 for sources for which an extinction curve is assigned.  Parameter estimation with CIGALE is based on Bayesian Inference, assuming Gaussian uncertainties \citep[][and references therein]{Walcher2011}. The probability distributions of the prior parameters of the models are assumed to be flat and the probability of all models are computed and integrated over all model parameters (except the one to be derived) to build the probability distribution function (PDF)  of the posterior. The mean and standard deviation of the PDF yield the parameter estimate and its associated uncertainty.
 The  validity of the measure is also checked  by using our mock analysis (cf. Sect. 3.3.2). In Fig. \ref{mockT}, the exact values of these parameters  are compared to their estimations obtained  by fitting the simulated samples built with each extinction curve. Both parameters  are reasonably well retrieved. Large $A_{\rm V}$ values of low  statistical significance\footnote{ The statistical significance is defined as  the ratio of the estimated value by  its  1 $\sigma$ uncertainty, both given by X-CIGALE} are underestimated. The dispersion of the estimated temperature corresponding to a true value of  100 K is large and its statistical significance is  low,  but similar for both extinction curves. So we expect that the difference found below between both distributions  is real.

\paragraph{Dust attenuation and X-ray absorbing column density}

In Fig. \ref{AV}, the  obscuration in the V-band due to polar dust, $A_{\rm V}$, is compared to the X-ray absorbing column density  $N_{\rm H}$  measured by \citet{Liu2016}\footnote{$\rm N_H$ has been calculated by \citet{Liu2016},  by applying X-ray spectral modeling and stacking,  adopting the Bayesian X-ray Analysis software (BXA,\citet{Buchner2014}) to fit the X-ray spectra of individual sources.}  for sources described with either an SMC or a G04 law. The  $A_{\rm V}$ values  are calculated with the color excess $E(B-V)_{\rm pd}$ and $R_{\rm V}$ corresponding to either the SMC or the G04 curve (cf. Sect. 2.1). $A_{\rm V}$ are found  lower in average for the SMC curve with all values lower than 1 mag,  than for the G04 curve  for which values extend up to 1.6 mag. This difference is explained by the shapes of the extinction curves: with an SMC curve the extinction  in UV ($\sim 150$ nm) is four times higher than in  the V-band  when both  extinctions are similar with the G04 curve. Therefore, the  total amount of energy absorbed by dust  is more efficient with a steep extinction curve  than with a flat curve for a given $A_{\rm V}$.  
Here, we also plot the $A_{\rm V}$ values corresponding to the 24 sources better fitted with the GB07 law (Sect. 3.3.3), with their distribution overlapping the one found with SMC, which is as expected since this law corresponds to a lower $R_{\rm V}$  value.

We do not find any correlation between $A_{\rm V}$ and $N_{\rm H,}$ but our sample is restricted to BLAGN   with  a  narrow range of   column densities  ($N_{\rm H}<10^{21.5} \rm{cm^{-2}}$ for most of the sources). The 
$A_{\rm V}/N_{\rm H}$ ratios  are distributed above and below the standard Galactic value ($A_{\rm V}/N_{\rm H}= 5.3 10^{-22} \rm{mag\: cm^{-2}}$ from \citet{Bohlin1978} with a standard Galactic $R_{\rm V}=3.1$). All our sources with an X-ray column density larger than $10^{21.5} \rm{cm^{-2}}$ have $A_{\rm V}/N_{\rm H}$ ratios lower than Galactic. 34$\%$  of the sources with an SMC curve have higher $A_{\rm V}/N_{\rm H}$ than Galactic and the percentage reaches 80$\%$ with a G04 extinction law  but, as mentioned above, the comparison between the G04 and Galactic laws is difficult because of    $R_{\rm V}$ difference. We note that the uncertainty in $N_{\rm H}$ is also large, even without accounting for X-ray variability. Using the value corresponding to the 84th percentile of the distribution reduces the fraction of sources above Galactic to $7\%$ (resp. $30\%$) for the SMC (resp. G04) law. 

The correlation between optical extinction and X-ray absorption is well established from unabsorbed  type 1  AGN to optically obscured type 2 AGN  although with a large dispersion \citep[e.g.,][]{Goodrich1994, Burtscher2016}.  \citet{Jaffarian2020} found that the  large scatter in the relation they found between the color excess $E(B-V)$ and  $N_{\rm H}$ for BLAGN  could be explained by the variability of the X-ray column densities \citep[see also] []{Burtscher2016} . 

\citet{Maiolino2001b} reported $A_{\rm V}/N_{\rm H}$ values in average lower than Galactic in various classes of AGN with broad emission lines  and $N_{\rm H} >10^{21.5} \rm{cm^{-2}}$. The only  three sources in their sample with $ N_{\rm H} < 10^{21.5} \rm{cm^{-2}}$ have an $A_{\rm V}/N_{\rm H}$ ratio that is about four times higher than Galactic, but with low X-ray luminosity. \citet{Burtscher2016} measured $A_{\rm V}/N_{\rm H}$ for 25 local X-ray detected source  and found values close to or lower than   Galactic. The ratios for their sources with $ N_{\rm H} \le \sim 10^{21} \rm{cm^{-2}}$  (and corresponding to Seyfert type 1 to 1.5) are found consistent with the Galactic value, although they show that variability of $N_{\rm H}$  makes  the results uncertain. 

The physical context to interpret these observations is also extremely complex. Several processes are rivaling  to modify the size distributions of dust grains:  grain-grain collisions,  with shattering and coagulation, and grain growth (accretion) on gas phase metals \citep[e.g.,][]{Asano2013}. All of them have an impact on both the shape of the extinction curve and the dust to gas ratio for a similar initial dust composition and size distribution. \citet{Hirashita2012} showed that coagulation  flattens the extinction curve and grain growth increases the steepness of the extinction curve in the UV
and the amplitude of the bump at 217,5 nm for an initial Galactic dust composition. Accretion and coagulation  have different effects on the extinction cross section per H atom and consequently on the  $A_{\rm V}/N_{\rm H}$ ratio, which also depends on the initial  and final distributions of grains sizes \citep[e.g.,][]{Cardelli1989,Maiolino2001b}. The depletion  of small grains can occur in the vicinity of the AGN, small grains being expelled by strong winds \citep{Gaskell2004}. Last but not least, radiation transfer can make the situation even more complex. \citet{Scicluna2015} showed that radiation transfer effects inside an observing beam that  encompasses a  dusty medium may also alter the effective extinction curve without any change in dust grain composition and {\bf are able to  reproduce} any $R_{\rm V}$ values.  

\begin{figure}
 \centering
\includegraphics[width=8cm]{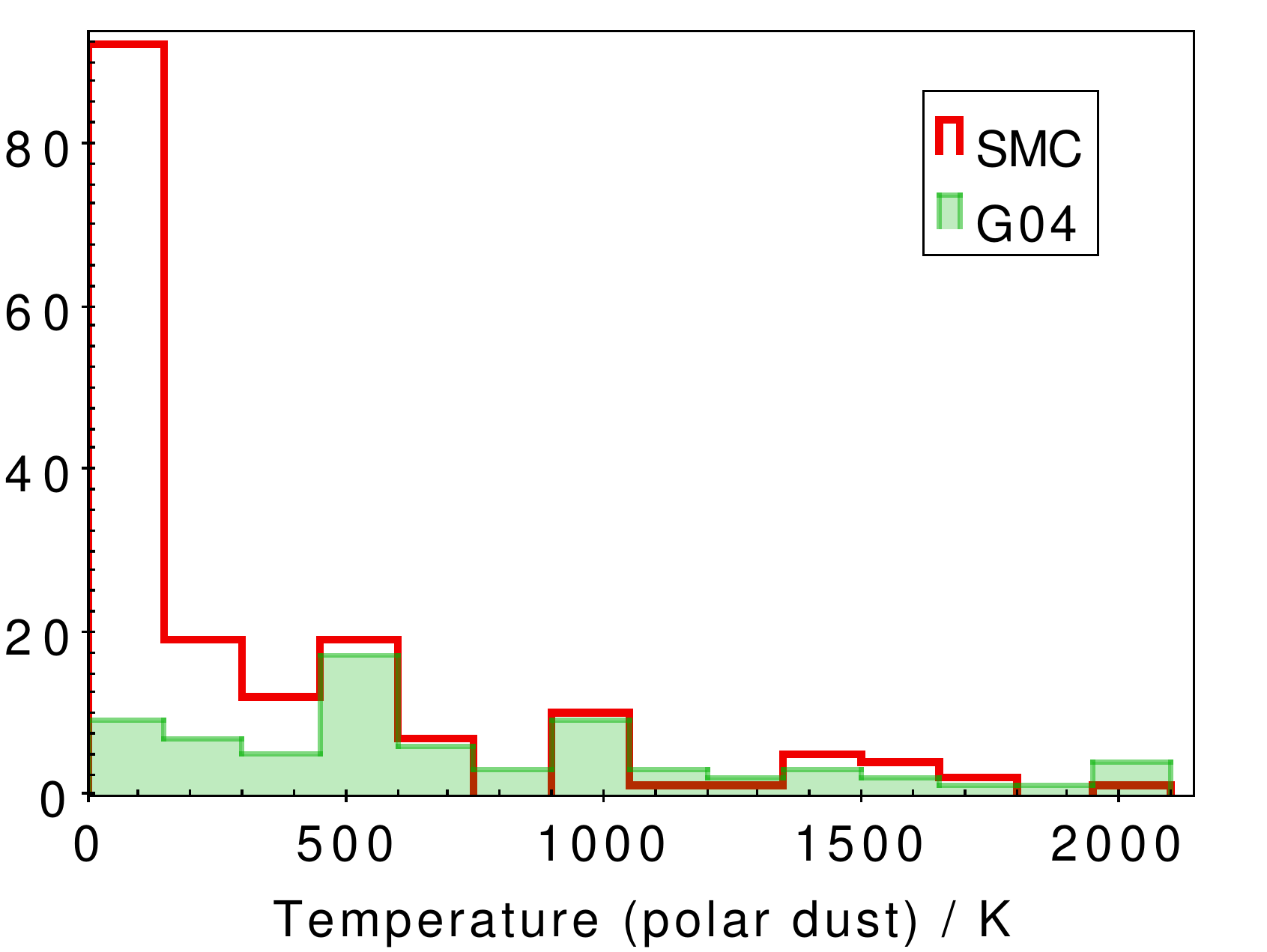}
\caption{Distribution of  polar dust temperatures of sources with polar dust described either with the    SMC law (red line) or the G04 law (green filled histogram). }
\label{Tpolar}
\end{figure}

\paragraph{Polar dust temperature}
The temperature of the  modified black body describing the emission of polar dust is a free parameter of our fitting process. The distributions of  values found for the two sub-samples described with  either an SMC or a G04 extinction curve are shown in Fig. \ref{Tpolar}.  Lower temperatures are found for polar dust better described with the SMC law than  with the G04  law.  The  temperatures  found with the G04 law extend over the full range of initial values (from 100 to 2000 K)  when 78$\%$ of the temperatures  for the SMC law  are   found lower than 500 K.  Very similar distributions are found when the sample is reduced to sources detected with {\it Herschel} as well as using the GB07 extinction curve  instead of G04 (cf. Sect. 3.3.4).
  The higher temperatures found with the G04 curve are likely to be  more easily found close to the AGN, but  they are not high enough for a destruction of small grains by  sublimation. 
 A  steep extinction curve, modeled with the  SMC law in this work,  could be explained by the presence of a  foreground, optical thin shell \citep{Witt2000, Scicluna2015}   located  at larger distance of the AGN  in consistency  with the lower temperatures  found for the SMC polar dust component.

 \section{Properties of BLAGN  with and without  polar dust}
 
 In this section, we compare the  distributions of  redshift, X-ray luminosity,  SFR, $\rm M_{star}$, and  specific SFR (defined as  $\rm sSFR = SFR/M_{star}$)  of  sources described with polar dust or without polar dust, as defined at the end of Sect. 3.3.1.  We restrict the analysis to sources with secure measurements of SFR and $\rm M_{star}$ as defined in Sect. 3.2. We are left with 301 sources with no polar dust and 289 sources with polar dust.  For sources with  polar dust but no clear assignment of an SMC or G04 extinction curve, we adopted the extinction curve giving the lowest $\chi^2$.   The estimated values of  SFR and  $\rm M_{star}$ are given in Table A.1.
 \begin{figure}[!]
   \centering
    \includegraphics[width=0.8\hsize,, angle=0]{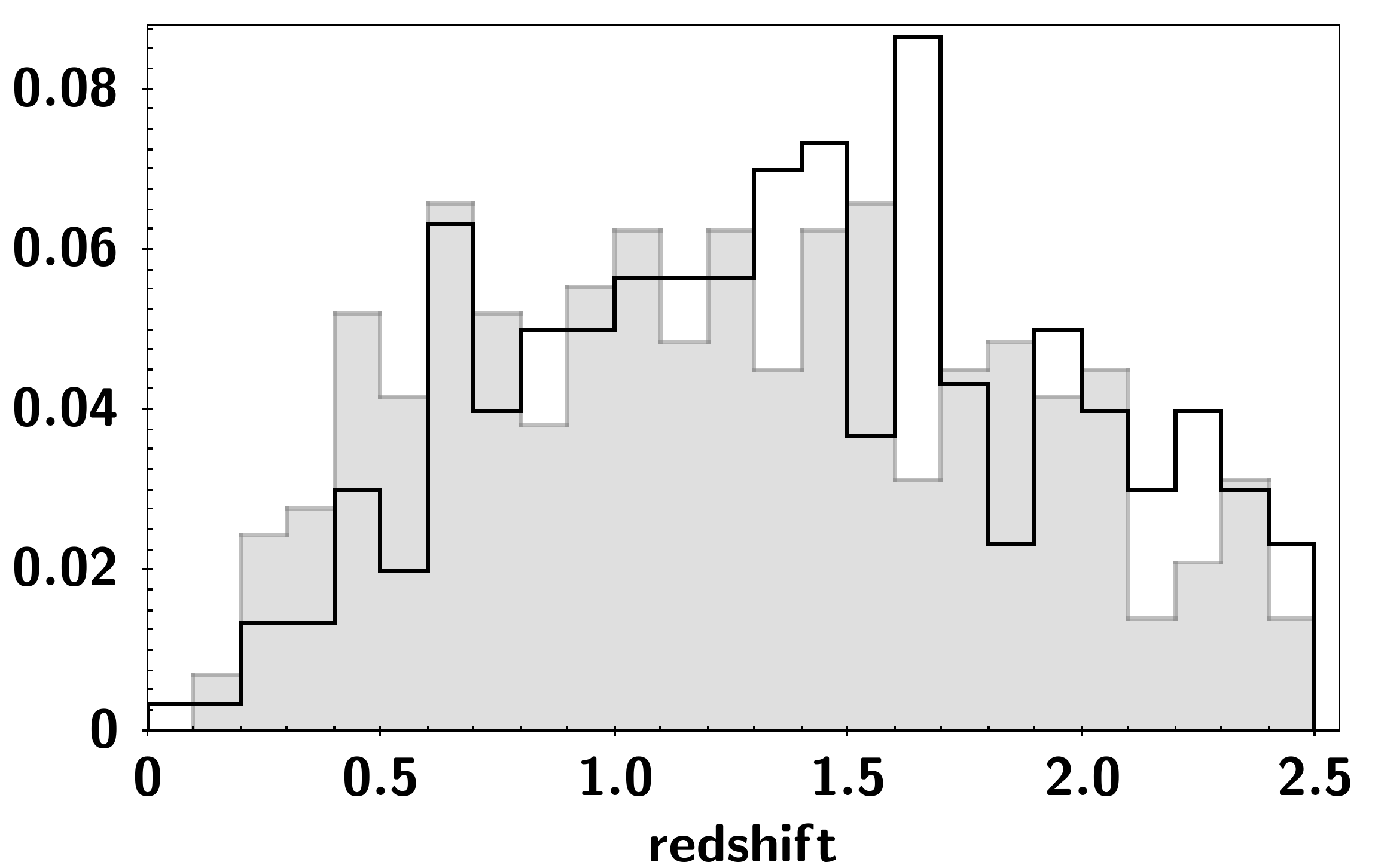}  
    \includegraphics[width=0.8\hsize,, angle=0]{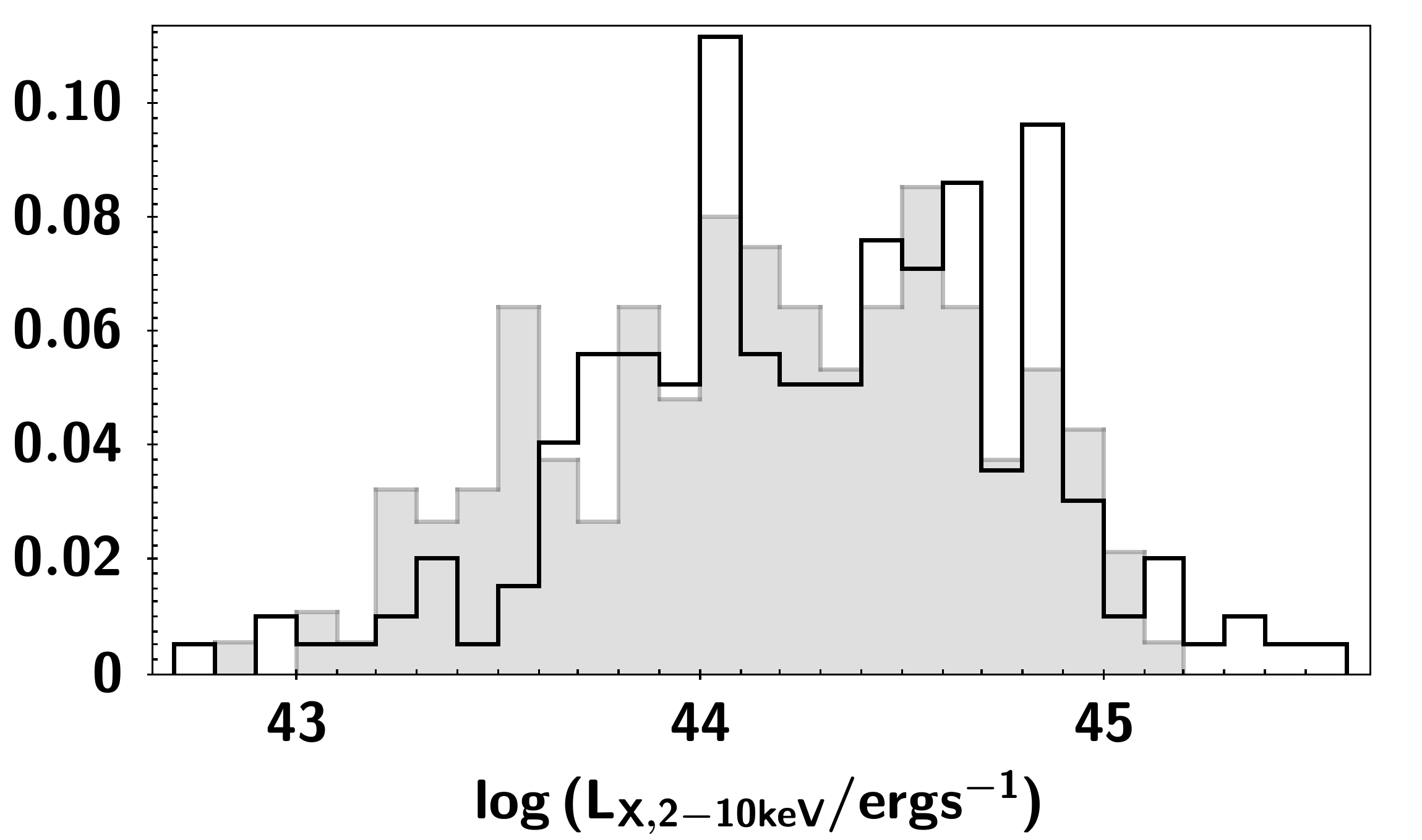}
         \caption{Distributions of redshift and X-ray luminosities for galaxies whose SED is better described with polar dust (filled histogram) and without polar dust (empty histogram).}
     \label{PD-noPD-z}
  \end{figure}  
  
   \begin{figure} 
       \centering              
     \includegraphics[width=0.8\hsize, angle=0]{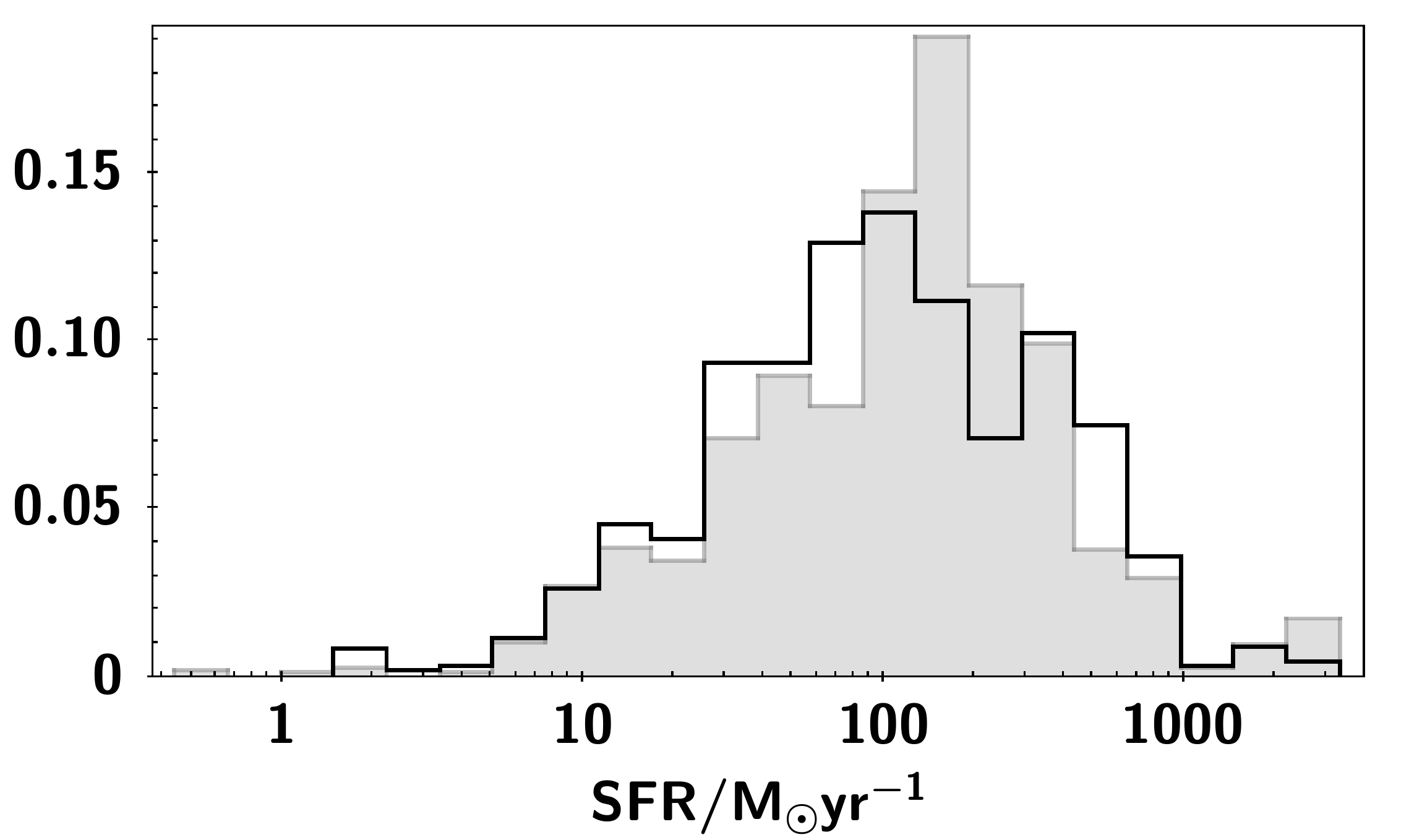}
        \includegraphics[width=0.8\hsize, angle=0]{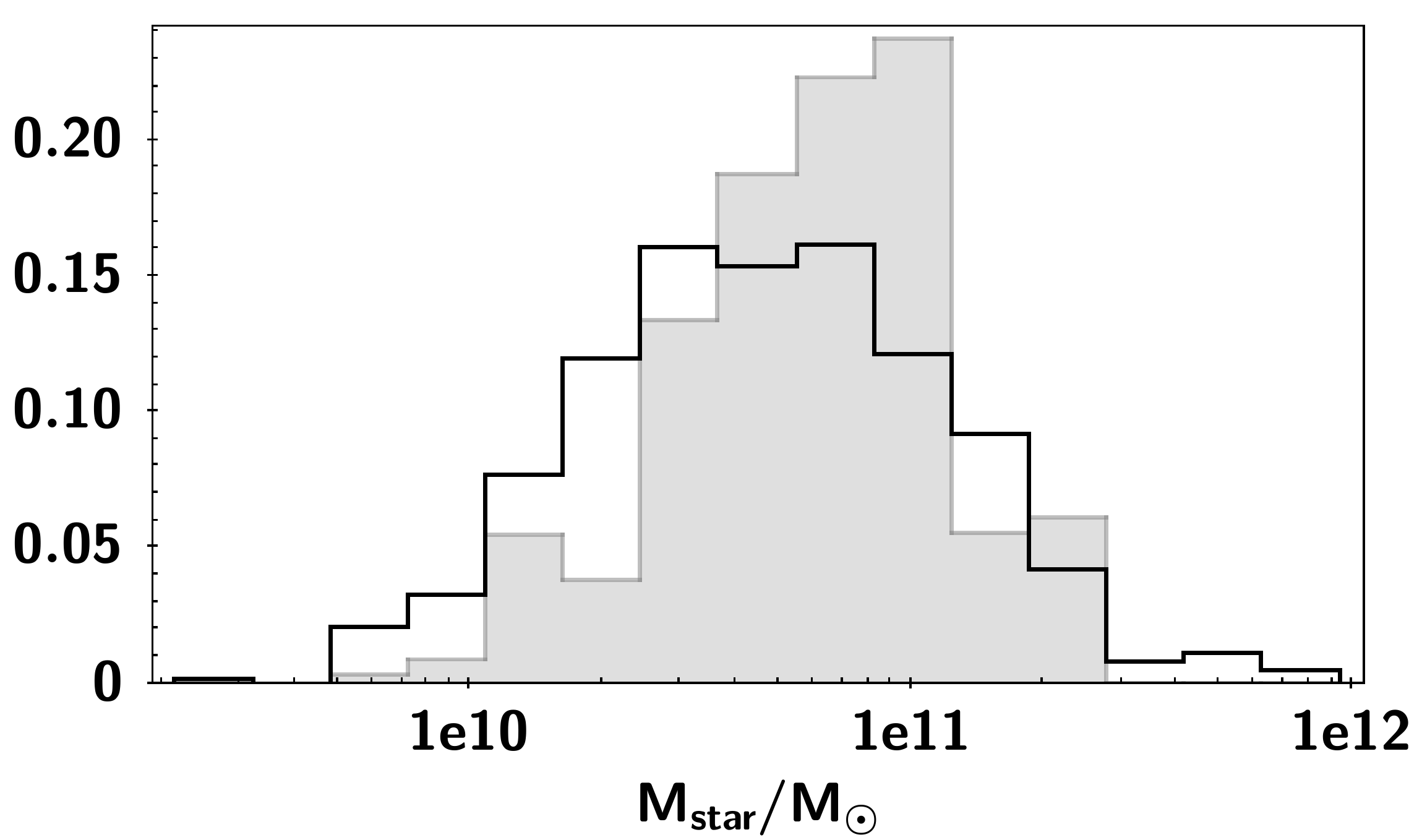}
         \includegraphics[width=0.8\hsize, angle=0]{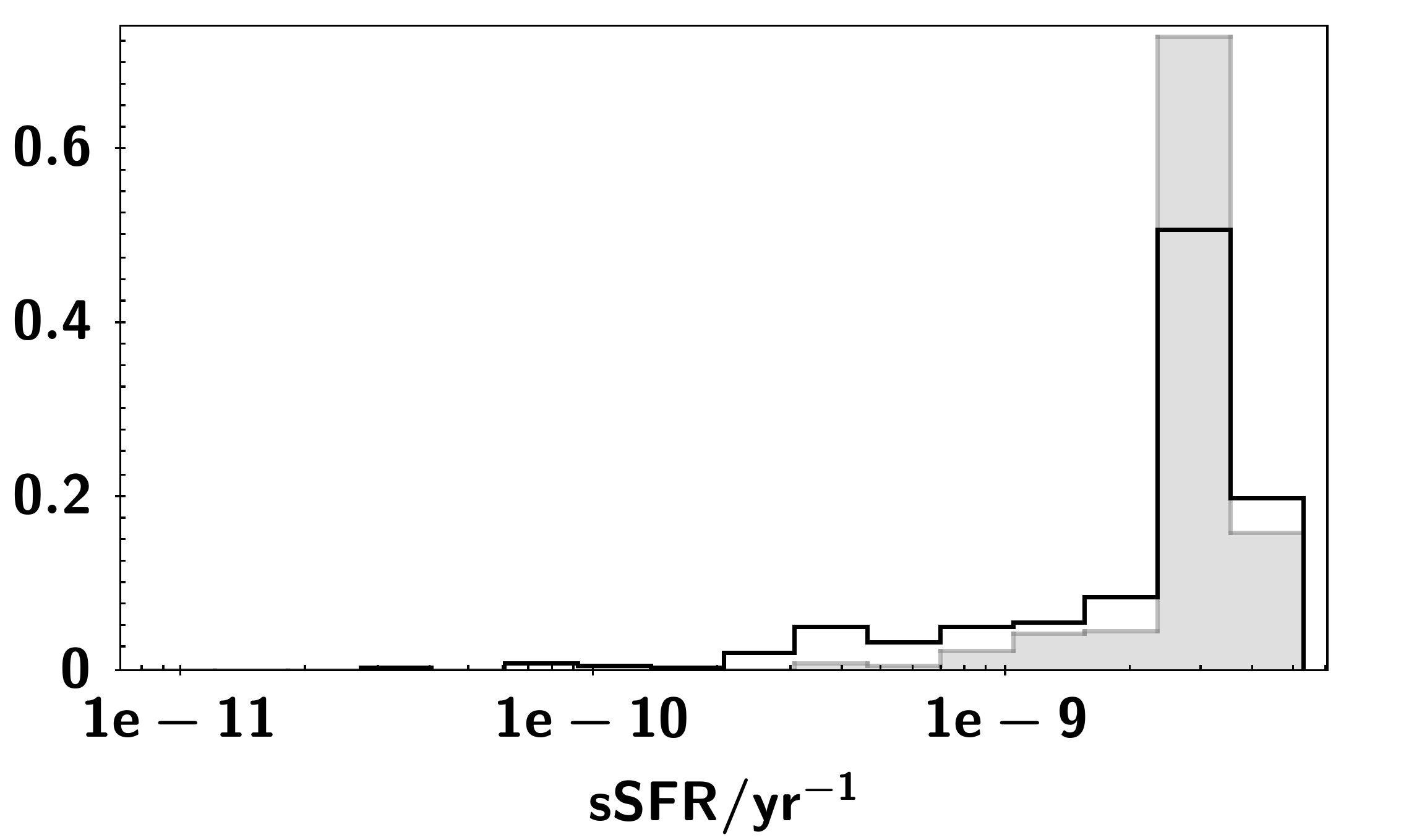}
       \caption{ Distributions of SFR,  $\rm M_{star}$  and  sSFR  for galaxies  with polar dust (filled histogram) and without polar dust (empty histogram).}
         \label{PD-noPD}
    \end{figure}  


The redshift distribution of  sources with and without polar dust  is shown in Fig. \ref{PD-noPD-z}. There is a small  excess of sources with polar dust at redshift lower than 1.  We performed a two-sample  Kolmogorov-Smirnov test and found a  p-value equal to 0.03.  We  account for this difference before comparing other parameters which may evolve with redshift 
 \citep[e.g.,][]{Masoura2018}. In each redshift bin, we randomly draw an equal number of sources with and without polar dust, the number of sources from each sub-sample being the lowest observed number in the bin. At the end of the process, we are left with 494 sources instead of the 590 initial ones. The distribution of the X-ray luminosity in the 2-10\,keV band\footnote{Hard X-ray luminosities  2-10\,keV  from the catalog of \citet{Menzel2016}.}  for the 494 sources,  is also  shown in Fig. \ref{PD-noPD-z}. A two-sample Kolmogorov-Smirnov test  returns a  p-value equal to 0.18 for the $\rm L_X$ distributions meaning that they cannot be distinguished.
  
We compare SFR, $\rm M_{star}$  and sSFR  for both sub-samples. These quantities are estimated with X-CIGALE with an uncertainty which depends on the SED of the source. We account for these uncertainties by weighting each measure with its statistical significance.  The resulting distributions are shown in Fig. \ref{PD-noPD}. For each considered quantity, the distributions  for sources with and without polar dust are  found significantly different from a two sample Kolmogorov-Smirnov test, the difference being  more pronounced for  $\rm M_{star}$ (p-value=0.0002 compared to  0.03 for SFR and sSFR). SFR and $\rm M_{star}$ are found shifted toward higher values for sources with polar dust, but mean  values of the SFR (resp.  $\rm M_{star}$) distributions differ by only 0.08 dex  (resp. 0.12 dex). The sSFR distribution of sources without polar dust shows a tail toward low values that is not found for the sources with polar dust, median values of the distributions differ by only 0.05 dex (average values are not considered because the sSFR distributions are  obviously not symmetrical).
 These differences in SFR and $\rm M_{star}$ distributions \citep[see also][]{Zou2019,Mountrichas2021} could be interpreted as an increase of dust mass in the sources with polar dust  according to well known scaling relations between dust masses,  $\rm M_{star}$,  and SFR  \citep[e.g.,][]{Cortese2012, Santini2014,dacunha2010}.  This could also serve as an argument in favor of  a  polar dust distributed over large scales and connected to the interstellar medium of the  galaxy. 
 In any case, these results have to be considered tentative and the small difference found between median or mean values of the distributions ($\sim 0.1$ dex)  is lower than the global expected uncertainty in $\rm M_{star}$  and SFR measurements \citep{Mountrichas2021b, Mountrichas2021}.

\begin{figure}[!]
   \centering
   \includegraphics[width=0.8\hsize, angle=0]{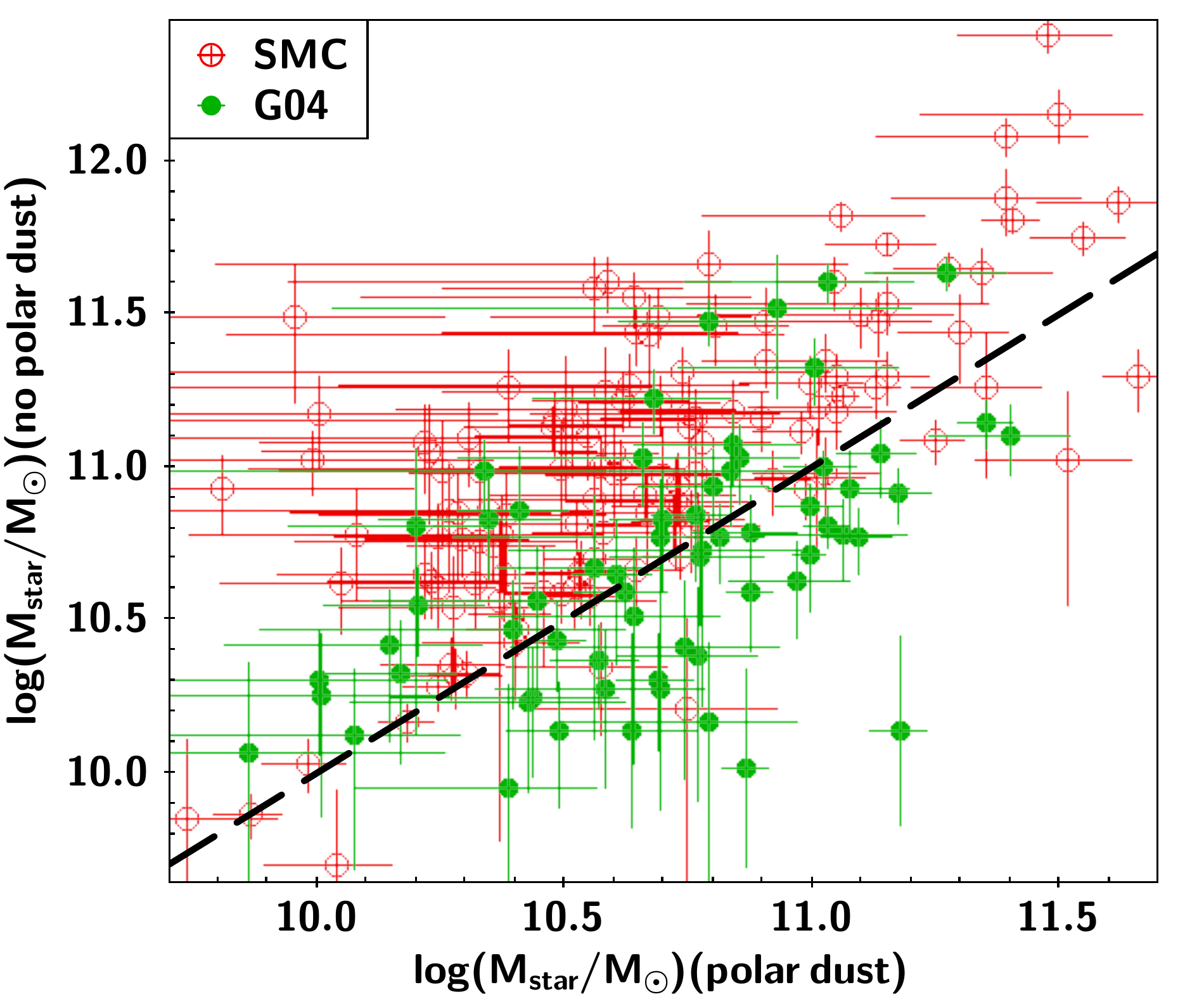}
  \includegraphics[width=0.78\hsize, angle=0]{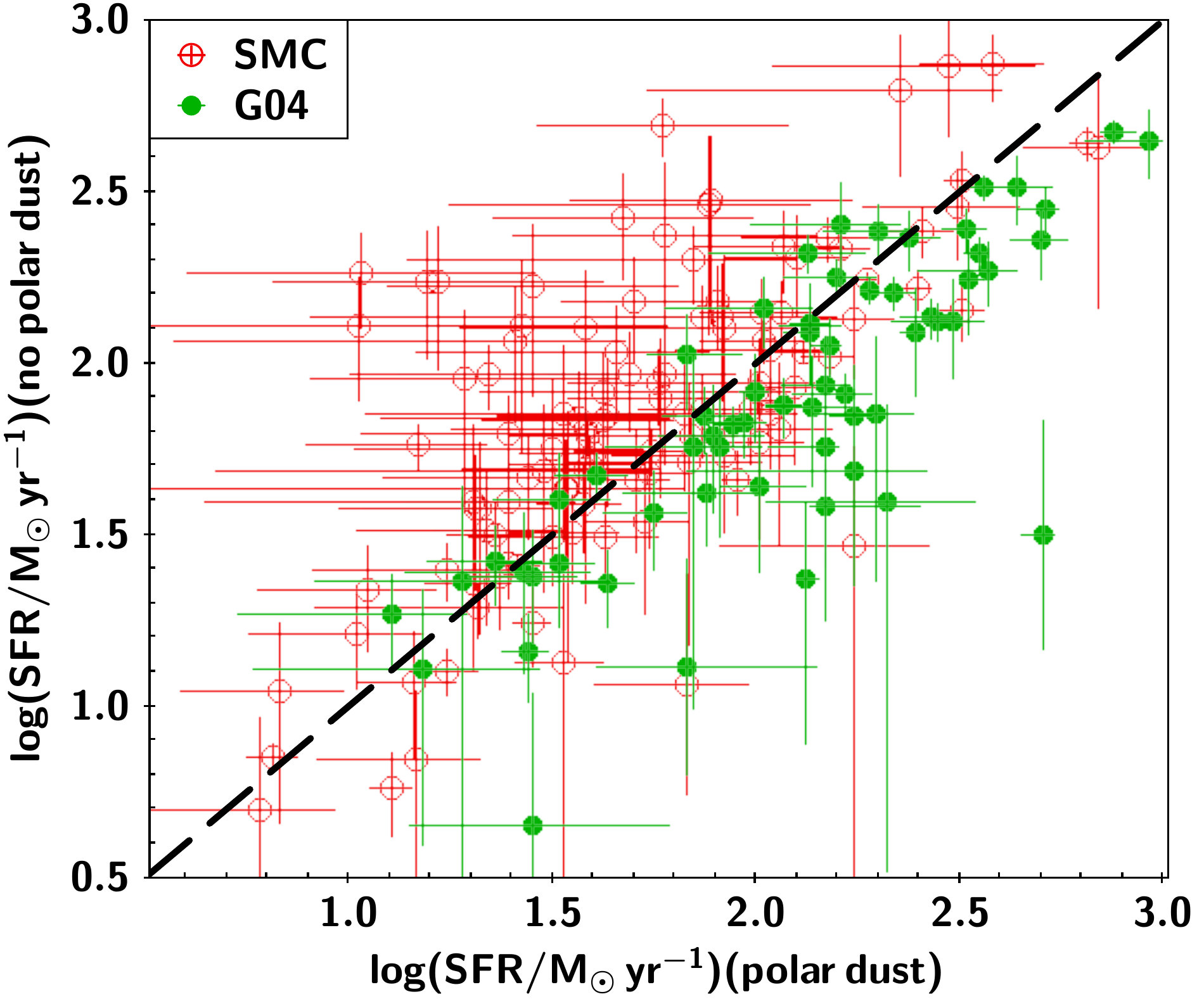}
       \caption{ SFR ({\it bottom panel}) and $\rm M_{star}$  ({\it top panel})  estimations   and their related uncertainties}  for sources  with polar dust (red dots: SMC law, green dots: G04 law). x-axis: values obtained  from the  fit with polar dust component,  y-axis: values obtained from the fit without polar dust component. 
         \label{PD-param}
   \end{figure}

\section{Impact of the polar dust component on the measurements of physical  parameters}

Adding a dust absorption to the emission of the active nucleus is expected to have an  impact on the contribution of the observed AGN emission to  the composite SED and, consequently, to the characteristics of the stellar component.
In Fig. \ref{PD-param}, we compare the estimated values  of  $\rm M_{star}$  and SFR for the sources  with polar dust, when the fit is performed with or without  polar dust.  When  polar dust is not considered, both quantities  are systematically overestimated  for sources best described with the SMC law. We have seen that the UV to near-IR spectra of the AGN component  is substantially reddened with  SMC polar dust (Sect. 2, Fig. \ref{models}, and Fig. \ref{extcurve}). When the composite SEDs are fitted without polar dust, the reddening   is only possible through  the obscuration of the stellar component: its contribution to the SED increases,  leading to higher $\rm M_{star}$   and SFR. When the polar dust is described with the G04 law, no systematic shift is found since this extinction law  does not imply any substantial reddening.

\begin{figure}[!]
  \centering
  \includegraphics[width=7cm]{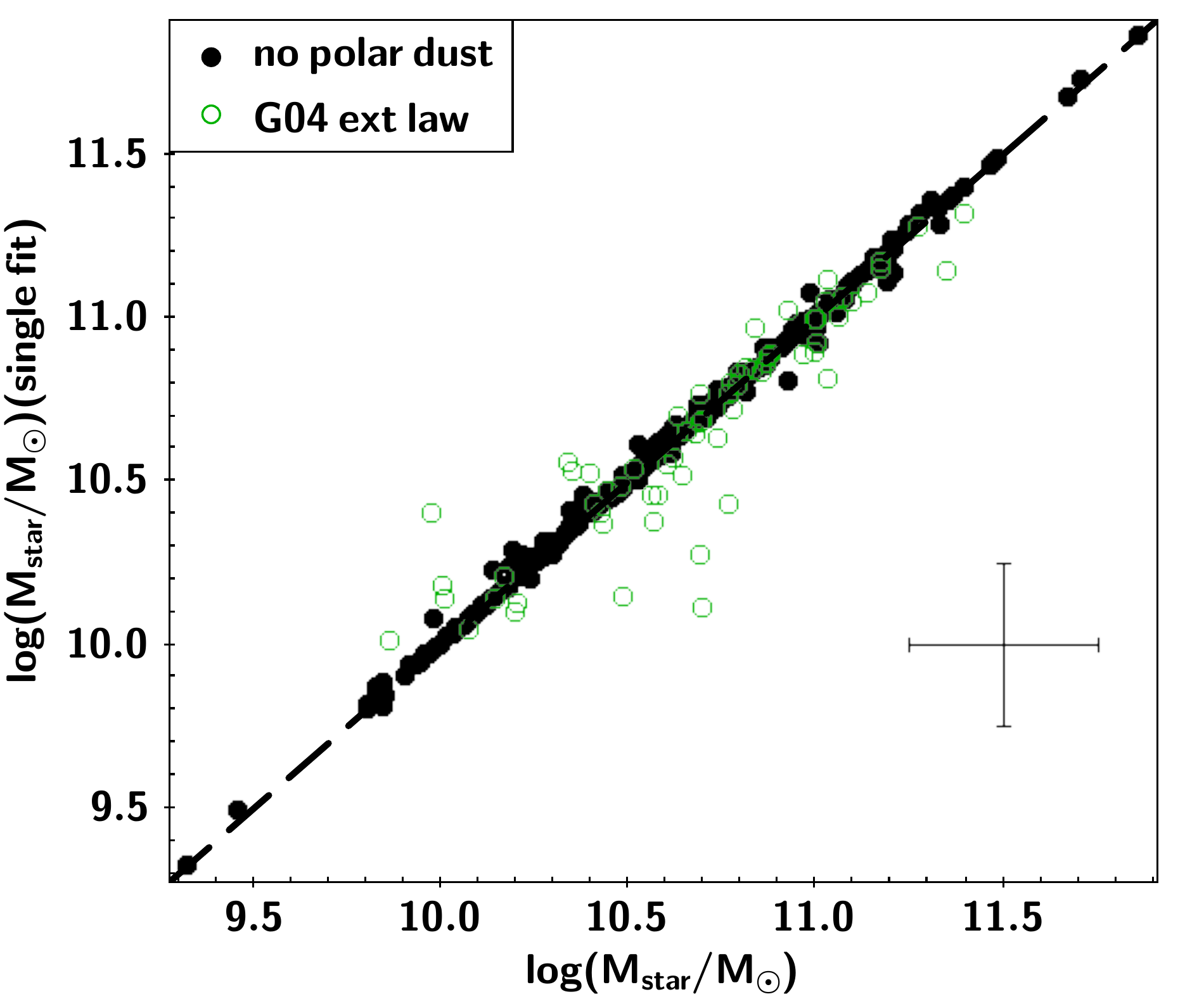}
  \includegraphics[width=7cm]{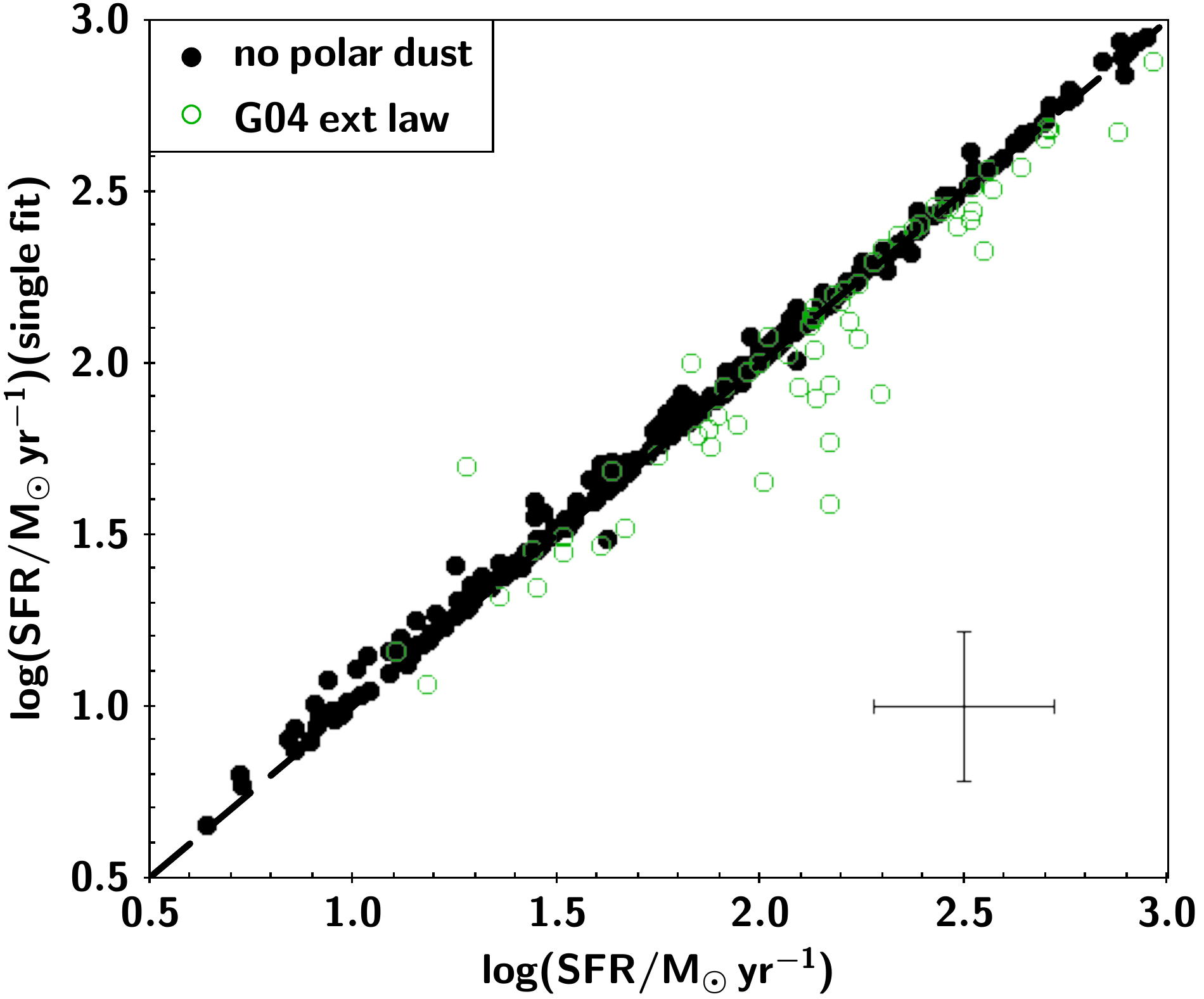}
    \caption{ SFR ({\it bottom panel}) and  $\rm M_{star}$ ({\it top panel})  estimated from a single fit assuming a polar dust obscuration with  an SMC extinction curve, for sources first identified with either polar dust described with a G04 curve (green empty circles) or no polar dust  (black dots). x-axis: Parameters estimated from the initial analysis (cf. Sect. 3), y-axis: Parameters estimated with the single fit.  The average 1$\sigma$ uncertainty on the estimated parameters is indicated as a cross.}
         \label{PD-param2}
   \end{figure}
   
The SEDs showing evidence against polar dust are selected to have a very low, even null, contribution of polar dust when this component is added 
(they are selected with  best fits having $E(B-V)_{\rm pd}=0$, and the likelihood weighted  values of $E(B-V)_{\rm pd}$ are all found lower than 0.05 mag). Therefore, performing  a fit with a polar dust component  includes  no polar dust  models. Parameter values derived from this fitting process  should  also be valid for  sources without polar dust: by construction, the values corresponding to the best fit will be identical  and only slight differences  are   expected for the likelihood weighted estimations of the parameters.\\
We now test the reliability of  results if only  a single fit of the SEDs of all our sources is performed.  It  is logical to  use  the SMC curve, since this curve is more often  preferred  than the G04 curve in our selection process and we showed that its introduction is mandatory to describe highly reddened sources.  In Fig. \ref{PD-param2}, the  SFR and $\rm M_{star}$  measurements with this single fit  are  compared  to SFR and $\rm M_{star}$  values obtained previously for sources with no polar dust or with  polar dust described with the G04 polar dust (cf. Sect. 3.2): the  agreement between both measurements is found  excellent for sources without polar dust  and both parameters are found  consistent   within  0.3 dex for more than 90$\%$ of the  sources with a polar dust  described with the  G04 curve from  our initial study.  We can conclude that a single fit accounting for a polar dust with the SMC extinction curve gives satisfactory results for measuring the   SFR and $\rm M_{star}$ of  host galaxies with a good accuracy.  However such a single fit only allows us to identify a polar dust component characterized by a steep extinction curve. In this work, we show  that the identification and the characterization of a  polar dust component  requires more  investigations based on  fits performed   with different extinction laws.

 \section{Summary}
In this work, we investigate the presence of  dust along the line of sight of   BLAGN by using the impact of adding  polar dust reddening to their UV to IR continuum. We fit the SEDs   of 1275 BLAGN  in the XMM-XXL field with X-CIGALE. The code  includes a simple model of polar component based on a screen geometry and an extinction curve, which can be chosen either steep (SMC curve of \citet{Pei1992}) or  almost flat in the UV-optical range (G04 extinction curve of  \citet{Gaskell2004}). Both curves have a very different impact on the reddening of the AGN continuum, which allows  us to compare  different configurations. 

We found a similar fraction  of sources that are better fitted  (or not)   with the introduction of a polar dust component. A mock analysis conducted to understand the performance of our classification led us to conclude that our sources  are  probably equally distributed between the two categories.
Among the sources identified with a polar dust, 50$\%$ are better modeled with an SMC curve against 21$\%$ with a G04 law. After corrections  from the mock analysis, we concluded that SMC-like curves could represent 60$\%$ of sources with polar dust against 40$\%$ for flat curves represented by the G04 law.  The fraction of sources found with or without polar dust is found nearly unchanged when the GB07 extinction  law  \citep{Gaskell2007} is used instead of G04, but the assignment of an extinction curve either flat or steep depends on the adopted curve. The GB07 extinction curve  is not as flat as G04 and makes the differentiation between  steep (SMC) and  flat (GB07) laws more difficult.

 The extinction in the V-band, $A_{\rm V}$, does not correlate with the X-ray absorbing column density $N_{\rm H}$. The distribution of $A_{\rm V}/N_{\rm H}$ ratios  is broad with values above and below Galactic. The  temperatures of the modified black body used for the dust polar re-emission extend over the full range of initial values (from 100 to 2000 K) for the G04  law, but 78$\%$ of the temperatures  for the SMC law were   found to be lower than 500 K.  These  results  could provide  some information on the properties of polar dust  and its location along the line of sight, although our very simple screen model on the one side and the complex physical processes assumed to be at work on the other side do not allow us to propose a fully consistent picture.
 
We showed that the fits of the SEDs without including polar dust lead to  an overestimation  of the stellar component (and, consequently, of SFR and $\rm M_{star}$) of sources with polar dust   described by the SMC law. A single fit  with X-CIGALE  that includes  a polar dust component   with the SMC extinction curve avoids this issue and returns reliable SFR and $\rm M_{star}$, but this is not sufficient to identify all the sources hosting polar dust. 

\begin{acknowledgements}
The project has received funding from Excellence Initiative of Aix-Marseille University - AMIDEX, a French 'Investissements d'Avenir' programme. GM acknowledges support by the Agencia Estatal de Investigac\'ion, Unidad de Excelencia Mar\'ia de Maeztu, ref. MDM-2017-0765. M.S. acknowledges support by the Ministry of Education, Science and Technological Development of the Republic of Serbia through the contract no. 451-03-9/2021-14/200002 and by the Science Fund of the Republic of Serbia, PROMIS 6060916, BOWIE.

\end{acknowledgements}

\nocite{*}
\bibliographystyle{aa}
\bibliography{aa-3}

\begin{appendix}
\onecolumn
\section{XCIGALE results for individual sources}
\begin{table*}[h]
\tiny
 \centering
 \caption{Properties derived  for each of the  1212 sources with good XCIGALE fits. SDSS-ID from SDSS-DR8 and  spectroscopic redshifts from \citet{Menzel2016} are listed in  columns 1 and 2. Polar dust prescription from Section 3.3.1 is summarized in column 3:  pd (nopd) is for sources for which polar dust (no polar dust) is needed, when an extinction curve is preferred it is indicated as SMC or G04, no indication means that nothing is concluded on the presence or not of polar dust. Polar dust temperature and  extinction in the V-band  and their 1$\sigma$ uncertainties (Sect. 3.3.3) are listed in columns 4 to 7. SFR and   $M_{\rm star}$  of sources found with or without polar dust (pd/nopd) (Section 4) are listed in columns 8 to 11 with  their 1$\sigma$ uncertainties. The full  table is available online.}
 \label{Tab.Table}
\setlength{\tabcolsep}{0pt}

\sisetup{
         table-column-width=13ex,    
         detect-all,
         round-mode=places,
         round-precision=2,
         tight-spacing,
         table-format = 4.2e-2,
         table-number-alignment = center
         }
\tiny
\begin{tabular}{|c|S|c|S|S|S|S|S|S|S|S|}
 
 \hline
 \hline
  \multicolumn{1}{|c|}{SDSS-ID} &
  \multicolumn{1}{c|}{redshift} &
  \multicolumn{1}{c|}{polar dust} &
  \multicolumn{1}{c|}{T$_{\rm pd }$} &
  \multicolumn{1}{c|}{T$_{\rm pd}$\_err} &
  \multicolumn{1}{c|}{A$_{\rm V}$} &
  \multicolumn{1}{c|}{A$_{\rm V}$\_err} &
  \multicolumn{1}{c|}{SFR} &
  \multicolumn{1}{c|}{SFR\_err} &
  \multicolumn{1}{c|}{M$_{star}$} &
  \multicolumn{1}{c|}{M$_{star}$\_err} \\

\multicolumn{1}{|c|}{} &
  \multicolumn{1}{c|}{} &
  \multicolumn{1}{c|}{} &
  \multicolumn{1}{c|}{K} &
  \multicolumn{1}{c|}{K} &
  \multicolumn{1}{c|}{mag} &
  \multicolumn{1}{c|}{mag} &
  \multicolumn{1}{c|}{${\rm M_\odot yr^{-1}}$} &
  \multicolumn{1}{c|}{${\rm M_\odot yr^{-1}}$} &
  \multicolumn{1}{c|}{${\rm M_\odot}$} &
  \multicolumn{1}{c|}{${\rm M_\odot}$} \\
\hline
  1237669769764536453 & 1.42799997 & nopd &  &  &  &  & 109.3110927826472 & 63.94456622220792 & 4.402360795172176E10 & 2.639949608915235E10\\
  1237669769764536641 & 0.569500029 & - &  &  &  &  &  &  &  & \\
  1237669769764864225 & 1.61220002 & nopd &  &  &  &  &  &  &  & \\
  1237669769764864528 & 2.171 & pd &  &  &  &  & 123.53527268083545 & 71.40338844320466 & 1.0431936454065785E11 & 5.4195479496000694E10\\
  1237669769764929773 & 1.34360003 & - &  &  &  &  &  &  &  & \\
  1237669769764995194 & 1.34210002 & nopd &  &  &  &  & 123.83981935144051 & 30.065592706405727 & 4.173397988264547E10 & 1.4084370940680311E10\\
  1237669769764995335 & 1.65989995 & - &  &  &  &  &  &  &  & \\
  1237669769765060854 & 1.43470001 & - &  &  &  &  &  &  &  & \\
  1237669769765060892 & 1.42320001 & pd &  &  &  &  & 30.91531984804364 & 24.327112684112045 & 1.3944877117278267E10 & 1.03296333521961E10\\
  1237669769765060899 & 1.42610002 & - &  &  &  &  &  &  &  & \\
  1237669769765126151 & 0.986400008 & - &  &  &  &  &  &  &  & \\
  1237669769765126429 & 1.50779998 & - &  &  &  &  &  &  &  & \\
  1237669769765126432 & 2.00149989 & - &  &  &  &  &  &  &  & \\
  1237669769765126463 & 1.29809999 & SMC & 500.1594188506493 & 15.490014504255475 & 0.38873756485268357 & 0.1999460936521404 & 35.32364604104634 & 10.303251216718717 & 2.506439106215359E10 & 8.822876017484226E9\\
  1237669769765126491 & 2.0006001 & - &  &  &  &  &  &  &  & \\
  1237669769765126588 & 1.88230002 & SMC & 100.12849516083298 & 10.816165995464798 & 0.16589989143499478 & 0.07317878806660323 &  &  &  & \\
  1237669769765191910 & 1.38030005 & nopd &  &  &  &  & 133.75812065621642 & 17.192582032505864 & 4.348803183162287E10 & 7.949022330535247E9\\
  1237669769765257492 & 0.818499982 & - &  &  &  &  &  &  &  & \\
  1237669769765257512 & 1.34169996 & - &  &  &  &  &  &  &  & \\
  1237669769765388719 & 0.574199975 & pd &  &  &  &  & 3.8167589848484584 & 5.925931433300154 & 2.838975960573726E11 & 8.034552260717075E10\\
  1237669770301407391 & 1.22950006 & - &  &  &  &  &  &  &  & \\
  1237669770301407420 & 1.01400006 & - &  &  &  &  &  &  &  & \\
  1237669770301407431 & 1.25039995 & nopd &  &  &  &  &  &  &  & \\
  1237669770301472956 & 1.81169999 & SMC & 407.79716434699236 & 388.3610106289369 & 0.14690534991624318 & 0.02201993908653619 & 28.219822316980057 & 35.68408331773355 & 1.6914908097682983E10 & 1.980953848903853E10\\
  1237669770301604149 & 1.24230003 & - &  &  &  &  &  &  &  & \\
  1237669770301669622 & 1.56860006 & - &  &  &  &  &  &  &  & \\
  1237669770301669638 & 1.64499998 & - &  &  &  &  &  &  &  & \\
  1237669770301735049 & 1.64230001 & nopd &  &  &  &  & 175.66240902726938 & 64.63735414290488 & 5.312943180070982E10 & 2.05043784885993E10\\
  1237669770301735095 & 2.23900008 & - &  &  &  &  &  &  &  & \\
  1237669770301800647 & 1.18620002 & SMC & 258.9704389800217 & 334.48281681855053 & 0.27874312670603973 & 0.07637376014620603 &  &  &  & \\
  1237669770301800653 & 1.18620002 & nopd &  &  &  &  & 9.546282392779 & 12.875835816719288 & 6.585733876342652E9 & 8.066100967850573E9\\
  1237669770301800672 & 1.11500001 & nopd &  &  &  &  &  &  &  & \\
  1237669770301800720 & 0.559000015 & - &  &  &  &  &  &  &  & \\
  1237669770301801194 & 0.671000004 & SMC & 1057.24136759252 & 666.6608789856805 & 0.4916311345825405 & 0.21522978672572804 & 20.440130112731 & 10.940575635731854 & 2.0840690581883152E10 & 6.81740763825959E9\\
  1237669770301866234 & 2.42650008 & nopd &  &  &  &  & 83.04988463792381 & 75.08715193394025 & 4.822288509259372E10 & 4.070167514855429E10\\
  1237669770301866271 & 1.48800004 & nopd &  &  &  &  &  &  &  & \\
  1237669770301866306 & 0.784399986 & - &  &  &  &  &  &  &  & \\
  1237669770301931831 & 2.04789996 & nopd &  &  &  &  & 94.69225098525376 & 78.92889290700883 & 3.3632731472779335E10 & 2.751221044749248E10\\
  1237669770838278217 & 1.08290005 & pd &  &  &  &  & 84.38523237205716 & 35.514584620225826 & 2.697970668557288E10 & 1.220407321768177E10\\
  1237669770838278404 & 1.51100004 & nopd &  &  &  &  & 101.60847929309868 & 24.40001679292839 & 3.1966639929424263E10 & 9.464846459819397E9\\
  1237669770838343941 & 0.703199983 & - &  &  &  &  &  &  &  & \\
  1237669770838343955 & 1.00090003 & - &  &  &  &  &  &  &  & \\
  1237669770838540577 & 1.54170001 & - &  &  &  &  &  &  &  & \\
  1237669770838606090 & 1.53649998 & SMC & 103.565738412531 & 42.601395951749915 & 0.1605449159663295 & 0.056069751423377626 & 84.12768778443383 & 41.09899793283868 & 5.0143382591058E10 & 2.1053612082543766E10\\
   \hline
\end{tabular}
\end{table*}
\end{appendix}
\end{document}